\documentclass[journal]{IEEEtran}
\renewcommand{\arraystretch}{0.9}
\usepackage{amsfonts}
\usepackage{amssymb}
\usepackage{amsthm}
\usepackage{amsmath,amsfonts,amssymb}
\usepackage[dvips]{graphicx}
\usepackage{subfigure}
\usepackage{verbatim}
\usepackage{setspace}
\usepackage{bm}
\usepackage{algorithmic} 
\usepackage[ruled,vlined]{algorithm2e}
\usepackage{cite}

\usepackage{changepage}
\usepackage{pdfpages}
\usepackage{color}
\usepackage{epstopdf}
\usepackage{hyperref}
\usepackage{multirow}
\usepackage{makecell,array}
\usepackage{graphicx}

\allowdisplaybreaks

\IEEEoverridecommandlockouts

\usepackage{threeparttable}
\usepackage{diagbox}
\usepackage{tabularx}
\newcolumntype{C}{>{\centering\arraybackslash}X}

\renewcommand{\baselinestretch}{1.0}

\begin{document}
	
	\title{\huge A Survey on Reconfigurable and Movable Antennas for Wireless Communications and Sensing}
	\author{Wenyan Ma,~\IEEEmembership{Graduate Student Member,~IEEE,}
		Lipeng Zhu, ~\IEEEmembership{Member,~IEEE,}
		Yanhua Tan,	
		Beixiong Zheng,~\IEEEmembership{Senior Member,~IEEE,}
		Yujie Zhang,~\IEEEmembership{Member,~IEEE,}
		Yuchen Zhang,~\IEEEmembership{Member,~IEEE,}
		Keke Ying,
		Zhen Gao,~\IEEEmembership{Member,~IEEE,}
		He Sun,~\IEEEmembership{Member,~IEEE,}
		Xiaodan Shao,~\IEEEmembership{Member,~IEEE,}
		Zhenyu Xiao,~\IEEEmembership{Senior Member,~IEEE,}
		Dusit Niyato,~\IEEEmembership{Fellow,~IEEE,}
		and Rui Zhang,~\IEEEmembership{Fellow,~IEEE}
		\thanks{W. Ma, L. Zhu, H. Sun, and R. Zhang are with the Department of Electrical and Computer Engineering, National University of Singapore, Singapore 117583 (e-mail: wenyan@u.nus.edu, lipzhu@outlook.com, sunele@nus.edu.sg, elezhang@nus.edu.sg).}
		\thanks{Y. Tan and B. Zheng are with the School of Microelectronics, South China University of Technology, Guangzhou 511442, China (e-mail: tanyanhua06@163.com, bxzheng@scut.edu.cn).}
		\thanks{Y. Zhang is with the School of Electrical and Electronic Engineering, Nanyang Technological University, Singapore 639798 (e-mail: yujie.zhang@ntu.edu.sg).}
		\thanks{Y. Zhang is with the Electrical and Computer Engineering Program, Computer, Electrical and Mathematical Sciences and Engineering (CEMSE), King Abdullah University of Science and Technology (KAUST), Thuwal 23955-6900, Kingdom of Saudi Arabia (e-mail: yuchen.zhang@kaust.edu.sa).}
		\thanks{K. Ying and Z. Gao are with the School of Information and Electronics, Beijing Institute of Technology, Beijing 100081, China, also with the MIIT Key Laboratory of Complex-Field Intelligent Sensing, Beijing Institute of Technology, Beijing 100081, China, also with the Yangtze Delta Region Academy of Beijing Institute of Technology (Jiaxing), Jiaxing 314019, China, and also with the Advanced Technology Research Institute of Beijing Institute of Technology, Jinan 250307, China (e-mail: ykk@bit.edu.cn, gaozhen16@bit.edu.cn).}
		\thanks{X. Shao is with the Department of Electrical and Computer Engineering, University of Waterloo, Waterloo, ON N2L 3G1, Canada (e-mail: x6shao@uwaterloo.ca).}
		\thanks{Z. Xiao is with the School of Electronic and Information Engineering, Beihang University, Beijing, China 100191. (e-mail: xiaozy@buaa.edu.cn).}
		\thanks{D. Niyato is with the College of Computing and Data Science, Nanyang Technological University, Singapore 639798 (e-mail: dniyato@ntu.edu.sg).}
	}
	
	\maketitle
	
	\begin{abstract}
		Reconfigurable antennas (RAs) and movable antennas (MAs) have been recognized as promising technologies to enhance the performance of wireless communication and sensing systems by introducing additional degrees of freedom (DoFs) in tuning antenna radiation and/or placement. This paradigm shift from conventional non-reconfigurable/movable antennas offers tremendous new opportunities for realizing multi-functional, more adaptive, and efficient next-generation wireless networks. In this paper, we provide a comprehensive survey on the fundamentals, architectures, and applications of these two emerging antenna technologies. First, we provide a chronological overview of the parallel historical development of both RA and MA technologies. Next, we review and classify the state-of-the-art hardware architectures for implementing RAs and MAs, followed by a detailed comparison of their distinct mechanisms, performance metrics, and functionalities. Subsequently, we focus on various applications of RAs and MAs in wireless communication systems, analyzing their respective performance advantages and key design considerations such as mode selection, movement optimization, and channel acquisition. We also explore the significant roles of RAs and MAs in advancing wireless sensing and integrated sensing and communication (ISAC). Furthermore, we present numerical performance comparisons to illustrate the distinct characteristics and complementary advantages of RA and MA systems. Finally, we outline key challenges and identify promising future research directions to inspire further innovations in this burgeoning field.
	\end{abstract}
	\begin{IEEEkeywords}
		Reconfigurable antenna (RA), movable antenna (MA), wireless communication, wireless sensing, integrated sensing and communication (ISAC), hardware architecture.
	\end{IEEEkeywords}
	
	\section{Introduction} 
	\label{sec:intro}
	
	\subsection{Background} 
	\label{subsec:intro_background}
	During the past few decades, wireless communication technology has been rapidly transforming, driven by increasing connectivity demands and novel applications. The foundations of multiple-input multiple-output (MIMO) technology, which underpin these advancements, were established in seminal works exploring its capacity and practical implementation \cite{telatar1999capacity, Paulraj2004Anover, Stuber2004broadb, goldsmith2005wireless, TseFundaWC}. Cellular networks evolved from voice-centric systems to the versatile fifth generation (5G), providing enhanced mobile broadband (eMBB), ultra-reliable low-latency communication (URLLC), and massive machine type communication (MTC) \cite{Shariatmadari2015MTCfor5G, Dowhuszko2016delayMTC, Beyene2017narrowIoT, Hsieh2018LTEmachine, Hyder2014SmartGrid, Ghena2019challenge, Xia2020EmgMTC,jadc1,jadc3}. A key 5G enabler is massive MIMO (mMIMO), deploying significantly more base station (BS) antennas (tens to hundreds) than prior generations. This facilitates precise beamforming, spatial multiplexing, and improved spectrum efficiency \cite{Larsson2014massive, Lu2014Anover}. Sixth generation (6G) networks, envisioned to be launched around 2030, target the capabilities of terabit-per-second (Tbps) rates, sub-millisecond latency, native artificial intelligence (AI), and integrated sensing and communication (ISAC) \cite{dang2020should, Saad2020Avision, gawas2015overview, wang2023road, jiang2021the, chowdhury20206g, ITU6G,liu2022survey}. Realizing these goals, particularly using high-bandwidth millimeter-wave (mmWave) \cite{ayach2014spatia, zeng2016millim, zhu2019millim} and terahertz (THz) frequencies \cite{Lin2016subarray, Akyildiz2022Terahertz}, likely requires even extremely large-scale MIMO (XL-MIMO) \cite{lu2024nearXL, Wang2024XLMIMO, you2024next} to counteract path loss and form ultra-narrow beams \cite{Ning2023THzbeam}. Parallel evolution occurs in wireless local area networks (WLANs). Standards like IEEE 802.11ac (Wi-Fi 5) \cite{ieee80211ac}, 802.11ax (Wi-Fi 6/6E) \cite{ieee80211ax}, and the emerging 802.11be (Wi-Fi 7) \cite{ieee80211be} boost throughput and efficiency, notably via multiuser MIMO (MU-MIMO) by leveraging more access point (AP) antennas for simultaneous device connectivity and interference mitigation in dense user settings. Overall, wireless communication evolution increasingly leverages more antennas to enhance capacity, coverage, and efficiency.
	
	Concurrently, wireless systems are increasingly used for sensing the surrounding environment \cite{wei2022toward,kim2024role}. Wireless sensing uses electromagnetic (EM) waves for applications spanning from far-field detection, localization, and tracking to near-field imaging and fine-grained environmental/physiological monitoring. Modern radar systems often employ phased arrays, using multiple antennas for electronic beam scanning \cite{mailloux2005phased, wirth2005radar}. The evolution to digital MIMO radar using multiple transmit and receive antennas marked a significant advance. MIMO radar greatly enhances spatial resolution, parameter identifiability, and interference suppression through waveform diversity and spatial diversity from equipping more antenna elements. Employing more antennas in sensing apertures is crucial for the high resolution required by applications like autonomous driving and advanced surveillance \cite{10422881}. Furthermore, exploring higher frequencies (mmWave and THz) beyond traditional microwave bands offers larger bandwidths for finer range resolution. However, higher frequencies often demand larger antenna arrays to form high-gain beams, overcome propagation losses, and achieve sufficient angular resolution despite smaller element sizes. Thus, similar to communication systems, advanced wireless sensing increasingly relies on deploying more antennas and associated radio frequency (RF) chains.
	
	The evolution from 5G mMIMO to the anticipated requirements of 6G necessitates further exploitation of the spatial domain, as time-frequency resources are limited. This thus motivates the development of XL-MIMO systems to counteract path loss and form ultra-narrow beams, particularly at higher frequencies \cite{lu2024nearXL, Wang2024XLMIMO, you2024next}. While the deployment of such XL-MIMO systems enhances communication and sensing  performance, it also amplifies critical system-level challenges:
	\begin{itemize}
		\item \textbf{\textit{Signal Processing Overhead:}} While large arrays theoretically boost spectral efficiency via spatial multiplexing and/or beamforming, practical gains are often limited by overhead from the increasingly high-dimensional signal processing. Accurate channel state information (CSI) acquisition for massive antennas demands substantial pilot/feedback signaling, consuming time-frequency blocks and reducing network spectral efficiency \cite{Larsson2014massive, Lu2014Anover}. In multi-cell systems, pilot contamination worsens with mMIMO, thereby hindering performance gain and interference mitigation. Furthermore, the complexity of high-dimensional resource allocation in the spatial domain can be prohibitive, potentially causing suboptimal spectrum utilization.
		\item \textbf{\textit{Energy Consumption:}} System energy consumption, particularly at the infrastructure (e.g., communication BSs and radars), scales significantly with the number of active antennas associated with power amplifiers \cite{telatar1999capacity, Paulraj2004Anover}. For fully-digital systems, each antenna typically needs a dedicated RF chain with power-hungry components. Consequently, RF front-end power consumption increases linearly with the number of antennas. Moreover, real-time baseband processing energy for channel estimation and precoding/combining escalates dramatically, requiring powerful and energy-intensive processors. Managing the thermal load from numerous active components also increases energy demands of cooling systems.
		\item \textbf{\textit{Hardware Complexity and Cost:}} Integrating numerous antennas, RF chains, and interconnections into compact, practical form factors faces significant hurdles in size, weight, wiring, and manufacturing tolerances. Achieving high accuracy demands precise calibration of all antenna elements/RF chains, which tends to be significantly more complex and sensitive to temperature/aging drift with array size. Mutual coupling also becomes more pronounced and difficult to mitigate \cite{Wallace2004mutual, Chen2018mutual, Amani2022sparse}. Furthermore, the cumulative cost of numerous antennas, high-performance RF parts, fast data converters, powerful processors, and complex integration/testing can be prohibitive for future wireless applications.
	\end{itemize}

	Addressing these interconnected challenges in signal processing overhead, energy consumption, and hardware complexity is crucial for successfully deploying future large-array wireless systems.
	
	\subsection{Emerging Antenna Techniques and Motivation}
	\label{subsec:intro_existing}
	{To address the aforementioned challenges in large-scale antenna array-based wireless systems, several low-cost and energy-efficient antenna technologies have been developed. In particular, the high hardware complexity and power consumption associated with a large number of RF chains in mMIMO systems have motivated the development of various antenna selection (AS) strategies \cite{Molisch2004MIMOsys, Sanayei2004antennas}. These strategies dynamically choose a subset of the available antennas for operation, aiming to retain a significant portion of the mMIMO benefits while reducing the number of active antennas and the associated RF chains, thereby reducing cost and power consumption efficiently. In parallel, to reduce the number of active antennas while maintaining performance, thinned and sparse arrays have also been explored \cite{Haupt1994thinned, Cen2010sparse, greene1978sparse, roberts2011sparse, gazzah2009optimum}. Another approach to reduce hardware complexity is beamspace MIMO, which utilizes lens antenna arrays to concentrate signal energy and thus reduces the number of RF chains and phase shifters \cite{Brady2013beamspace, Zeng2014Lens, zeng2016millim}.}
	
	While these techniques have advanced the state-of-the-art of MIMO, they often rely on fixed antenna structures and configurations that cannot fully adapt to dynamic environments and/or diverse function requirements of contemporary communication and sensing systems. The limitations of existing approaches further motivate the exploration of more adaptable and intelligent antenna technologies. There is a pressing need for low-cost and versatile antenna solutions that can dynamically adjust their characteristics to optimize performance metrics such as capacity, reliability, energy efficiency, or sensing accuracy according to specific application requirements and prevailing channel conditions. This need drives the research into reconfigurable antennas (RAs) and movable antennas (MAs).

	\subsection{Introduction of RA and MA}
	\label{subsec:intro_rama_intro}
	To overcome the limitations of conventional non-reconfigurable/movable antennas and meet the demands of future wireless systems, RA and MA have emerged as promising paradigms offering enhanced degrees of freedom (DoFs) for system optimization \cite{zhu2025movablenpj}.
	
	\subsubsection{RA}
	\label{ssubsec:intro_ra}
	The RA primarily focuses on dynamically altering the antenna's internal operational characteristics \cite{costantine2015reconfigurable}. This reconfiguration can be achieved through various means, including electrical switching (e.g., using PIN diodes and micro-electromechanical system (MEMS) switches), mechanical adjustments (e.g., physical deformation), or by incorporating tunable materials (e.g., liquid crystals and ferrites) into the antenna structure \cite{Christodoulou2012reconfig, tandel2023reconfigurable}. By manipulating these internal mechanisms, RAs can dynamically adjust one or more of their fundamental properties, including
	\begin{itemize}
		\item \textit{Frequency Reconfiguration:} This capability allows an antenna to dynamically shift its operating frequency band or center frequency \cite{2018-TAP-FRE,2008-AWPL-FPRA,2009-AWPL-FrePRA,2022-TAP-FeRA,2006-TAP-FRe}. This provides versatility in multi-standard devices and allows for cognitive radio applications by selecting less congested frequency channels \cite{Filtenna, wideband_FRA}.
		\item \textit{Radiation Pattern Reconfiguration:} This involves altering the spatial distribution of the radiated power, effectively changing the antenna's beam shape or direction \cite{2024-TAP-PRA,2024-Eucap-Dirk,2023-OJAP-multi-portPRA}. Such adaptability enhances signal directivity, improves spatial filtering, and allows coverage area adjustments \cite{1978-Harrington}.
		\item \textit{Polarization Reconfiguration:} This refers to the ability to change the polarization state of the radiated EM wave, such as switching among linear vertical, linear horizontal, and circular polarizations (CPs) \cite{2020-TAP-pol.GuoYJ,2020-OJAP-pol,polar_mechan, polar_PIN}. Polarization agility helps mitigate polarization mismatch losses between transmitter and receiver, enhances diversity reception in fading environments, and can improve signal penetration or reduce specific types of interference.
	\end{itemize}

	Besides, a prominent subclass of RAs aims to maximize spatial DoFs through nearly continuous apertures. This leads to the concept of continuous-aperture MIMO (CAP-MIMO) \cite{Zhang2022pattern}, which is also called holographic MIMO (HMIMO) \cite{renzo2019smart,Pizzo2020spatial}, large intelligent surface \cite{Hu2018beyond}, or holographic surface \cite{Huang2020hologr, Wan2021HoloRIS}. By designing radiation apertures that are nearly continuous rather than composed of discrete elements, CAP-MIMO potentially offers the ultimate spatial resolution and beamforming capabilities. With recent advancements in metamaterials, intelligent reflecting surfaces (IRSs) or reconfigurable intelligent surfaces (RISs) \cite{liaskos2018new,wu2019IRS, Huang2019RIS, mei2022intelligent, zheng2022survey, wu2024ISfor6G, liu2021RIS} can be fabricated more cost-efficiently by integrating numerous low-cost passive elements. By controlling the phase shift imparted by each element, IRS can reshape the wireless propagation environment, reflecting incident signals towards desired receivers or away from non-intended receivers, thus enhancing communication links without introducing additional active antennas. Besides, dynamic metasurface antennas (DMAs) \cite{pulido2016application,Shlezinger2019DMA} and reconfigurable holographic surfaces (RHSs) employ metamaterial elements as active radiators \cite{deng2021RHS,Deng2021reconf}. These elements are closely packed to provide high aperture efficiencies and beamforming capabilities, eliminating the reliance on traditional phase shifters.

	RAs offer significant advantages over conventional non-reconfigurable antenna technologies by providing versatility and adaptability. Compared to large phased arrays, a single RA element or a small-size RA array can potentially achieve diverse functionalities (such as beam steering and frequency tuning) with significantly reduced hardware complexity, cost, and power consumption.

	\subsubsection{MA}
	\label{ssubsec:intro_ma}
	The MA introduces new spatial adaptability by dynamically altering the antenna's external properties, such as its physical position within a given space \cite{zhu2023MAMag,zhu2022MAmodel,ma2022MAmimo}. In general, the three-dimensional (3D) position and 3D orientation of antennas can be both adjusted, also known as (a.k.a) six-dimensional MA (6DMA) \cite{shao20246DMA, shao2024Mag6DMA, shao2024discrete}. This physical displacement is typically achieved using electromechanical systems such as motors, actuators, or MEMS integrated with the antenna element or sub-array. Specifically, 6DMA enables two categories of antenna reconfigurations, which are elaborated as follows \cite{shao20246DMA, shao2024Mag6DMA, shao2024discrete}.
	\begin{itemize}
		\item \textit{Position Reconfiguration (Translation):} This capability involves physically changing the antenna's location within a one-dimensional (1D)/two-dimensional (2D)/3D space by mounting the antenna directly on motor-driven shafts or using MEMS \cite{zhu2023MAMag}. This allows the antenna to dynamically seek locations with favorable channel conditions (e.g., stronger signal or less interference), effectively harnessing spatial channel variations for improving communication and sensing performance.
		\item \textit{Orientation Reconfiguration (Rotation):} This capability involves altering the antenna's pointing direction and/or boresight direction by changing its elevation or azimuth angle, without significantly changing its central position \cite{shao20246DMA, shao2024Mag6DMA, shao2024discrete,zheng2025rotatable,zhengtian2025rotatable}. Achieved through mechanical/electronic rotation or tilting mechanisms, it allows for fine-tuning the alignment of the antenna's main lobe towards a target or optimizing its polarization alignment relative to incoming signals. Orientation adjustments complement electronic beam steering and are crucial for optimizing links over large distances or in environments with specific angular requirements.
	\end{itemize}

	Generally speaking, MAs provide a different approach to spatial adaptation. By physically relocating, MAs can potentially access spatial DoFs unavailable to fixed antennas, leading to significant gains in signal strength, diversity, and interference mitigation, especially in environments with strong spatial variations \cite{zhu2022MAmodel, ma2022MAmimo}. MAs offer advantages over traditional AS by providing a continuous or finely discretized set of locations/orientations rather than choosing from a few fixed options \cite{zhu2022MAmodel, zhu2024nearfield, ding2024near}. While mechanical movement is typically slower than electronic reconfiguration of RAs, the potential performance gains derived from optimal positioning can be substantial, even for long-term performance optimized using  statistical channel knowledge \cite{chen2023joint,shao20246DMA,lll}.
	
	\begin{figure}[!t]
		\centering
		\includegraphics[width=80mm]{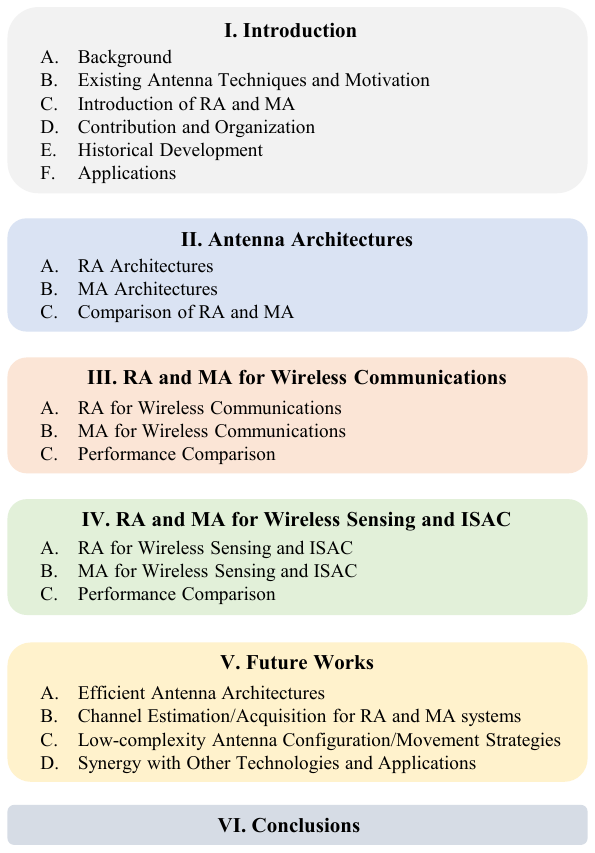}
		\caption{Organization of this paper.}
		\label{Fig_organization}
	\end{figure}

	\begin{table*}[]\footnotesize
		\centering
		\caption{List of acronyms.}
		\label{Tab_acronym}
		\begin{tabular}{|c|c|c|c|}
			\hline
			\textbf{Acronyms} & \textbf{Definition} & \textbf{Acronyms} & \textbf{Definition} \\ \hline
			1D & One-dimensional & LWA & Leaky-wave antenna \\ \hline
			2D & Two-dimensional & MA & Movable antenna \\ \hline
			3D & Three-dimensional & MEMS & Micro-electromechanical system \\ \hline
			5G & Fifth generation & MISO & Multiple-input single-output \\ \hline
			6G & Sixth generation & MIMO & Multiple-input multiple-output \\ \hline
			6DMA & Six-dimensional movable antenna & mMIMO & Massive multiple-input multiple-output \\ \hline
			AI & Artificial intelligence & MRA & Movable and reconfigurable antenna \\ \hline
			AirComp & Over-the-air computation & MSE & Mean square error \\ \hline
			AoA & Angle-of-arrival & MTC & Machine type communication \\ \hline
			AoD & Angle-of-departure & MU-MIMO & Multiuser multiple-input multiple-output \\ \hline
			AP & Access point & MUSIC & Multiple signal classification \\ \hline
			AS & Antenna selection & NEMS & Nano-electromechanical system \\ \hline
			BS & Base station & NGMA & Next-generation multiple access \\ \hline
			CAP-MIMO & Continuous-aperture MIMO & NLOS & Non-line-of-sight \\ \hline
			CDF & Cumulative distribution function & NOMA & Non-orthogonal multiple access \\ \hline
			CP & Circular polarization & OMP & Orthogonal matching pursuit \\ \hline
			CRB & Cramér-Rao bound & PCM & Phase change materials \\ \hline
			CRLH & Composite right/left-handed & PIFA & Planar inverted-F antenna \\ \hline
			CSI & Channel state information & PLS & Physical layer security \\ \hline
			DC & Direct current & PRM & Path response matrix \\ \hline
			DGS & Defected ground structures & PU & Primary user \\ \hline
			DMA & Dynamic metasurface antenna & RA & Reconfigurable antenna \\ \hline
			DoF & Degrees of freedom & RCS & Radar cross-section \\ \hline
			eMBB & Enhanced mobile broadband & RF & Radio frequency \\ \hline
			EM & Electromagnetic & RHCP & Right-hand circular polarization \\ \hline
			ESPAR & Electronically steerable parasitic array radiator & RIS & Reconfigurable intelligent surface \\ \hline
			FAS & Fluid Antenna Systems & RPA & Reconfigurable patch antenna \\ \hline
			FET & Field-effect transistor & RSMA & Rate-splitting multiple access \\ \hline
			FPA & Fixed position antenna & RSS & Received signal strength \\ \hline
			FPGA & Field-programmable gate array & SA & Sectorized antenna \\ \hline
			FRI & Field response information & SAR & Synthetic aperture radar \\ \hline
			FRV & Field response vector & SBA & Switched-beam antenna \\ \hline
			FSO & Free-space optical & SE & Spectral efficiency \\ \hline
			FSS & Frequency selective surface & SIMO & Single-input multiple-output \\ \hline
			GPS & Global positioning system & SINR & Signal-to-interference-plus-noise ratio \\ \hline
			HMIMO & Holographic MIMO & SISO & Single-input single-output \\ \hline
			i.i.d. & Independent and identically distributed & SNR & Signal-to-noise ratio \\ \hline
			IMS & Iterative mode search & SWIPT & Simultaneous wireless information and power transfer \\ \hline
			IRS & Intelligent reflecting surface & Tbps & Terabit-per-second \\ \hline
			ISAC & Integrated sensing and communication & THz & Terahertz \\ \hline
			LEO & Low-earth orbit & UAV & Unmanned aerial vehicle \\ \hline
			LHCP & Left-hand circular polarization & UE & User equipment \\ \hline
			LLM & Large language model & ULA & Uniform linear array \\ \hline
			LoS & Line-of-sight & UPA & Uniform planar array \\ \hline
			LP & Linear polarization & URLLC & Ultra-reliable low-latency communication \\ \hline
			VGA & Variable gain amplifier & WPT & Wireless power transfer \\ \hline
			WLAN & Wireless local area network & WSN & Wireless sensor network \\ \hline
			XL-MIMO & Extremely large-scale MIMO & ZF & Zero-forcing \\ \hline
		\end{tabular}
	\end{table*}

	\subsection{Contribution and Organization}
	\label{subsec:intro_contribution}
	While several review papers focus on RA \cite{Christodoulou2012reconfig, tandel2023reconfigurable} or, more recently, on MA/6DMA/FAS \cite{zhu2023MAMag, Shah2024survey, new2024tutorial, zheng2024flexible,zhu2025tutorial,shao2025tutorial}, this paper aims to provide a unified and comprehensive survey covering \textit{both} RA and MA/6DMA technologies within a single framework, as the two technologies share a similar principle in reshaping wireless channels in the EM domain. Given the significant technical potential and broad application prospects of RA and MA/6DMA, this paper provides a comprehensive survey on the fundamentals, architectures, and applications of these two promising antenna technologies, as well as their comparison. The main contributions of this paper are summarized as follows:
	\begin{itemize}
		\item We provide a comprehensive overview of the historical development of both RA and MA/6DMA technologies, tracing their parallel evolution within the antenna architecture, communication, and sensing communities.
		\item We review and classify the state-of-the-art hardware architectures for implementing RA and MA/6DMA, covering both element-level and array-level designs. A detailed comparison of their distinct mechanisms, performance metrics, and functionalities is also presented.
		\item We investigate the application of RA and MA/6DMA in wireless communication systems as well as analyze their respective performance advantages and key design considerations, including mode selection, movement optimization, and channel acquisition.
		\item We explore the significant roles of RA and MA/6DMA in advancing wireless sensing and ISAC, highlighting their unique benefits and design trade-offs.
		\item We present numerical performance comparisons to illustrate the distinct characteristics and complementary strengths of RA and MA/6DMA systems in various communication and sensing scenarios.
		\item We outline key challenges and identify promising future research directions to inspire further innovations in this burgeoning field.
	\end{itemize}

	The remainder of this paper is organized as follows. Section \ref{sec:architectures} provides a detailed review of antenna architectures for both RA and MA/6DMA, covering their classification, implementation methods, and design challenges, and concludes with a comparative analysis. Section \ref{sec:comms} focuses on the application of RA and MA/6DMA in wireless communications, examining their performance benefits and design methodologies. Section \ref{sec:sensing_isac} extends the discussion to wireless sensing and ISAC, reviewing RA- and MA/6DMA-enabled techniques and their integration. Section \ref{sec:future} outlines important future research directions. Finally, Section \ref{sec:conclusion} concludes the paper. The organization of this paper is summarized in Fig.~\ref{Fig_organization}. The acronyms used in this paper are summarized in Table \ref{Tab_acronym}.

	\begin{figure}[!t]
		\centering
		\includegraphics[width=88mm]{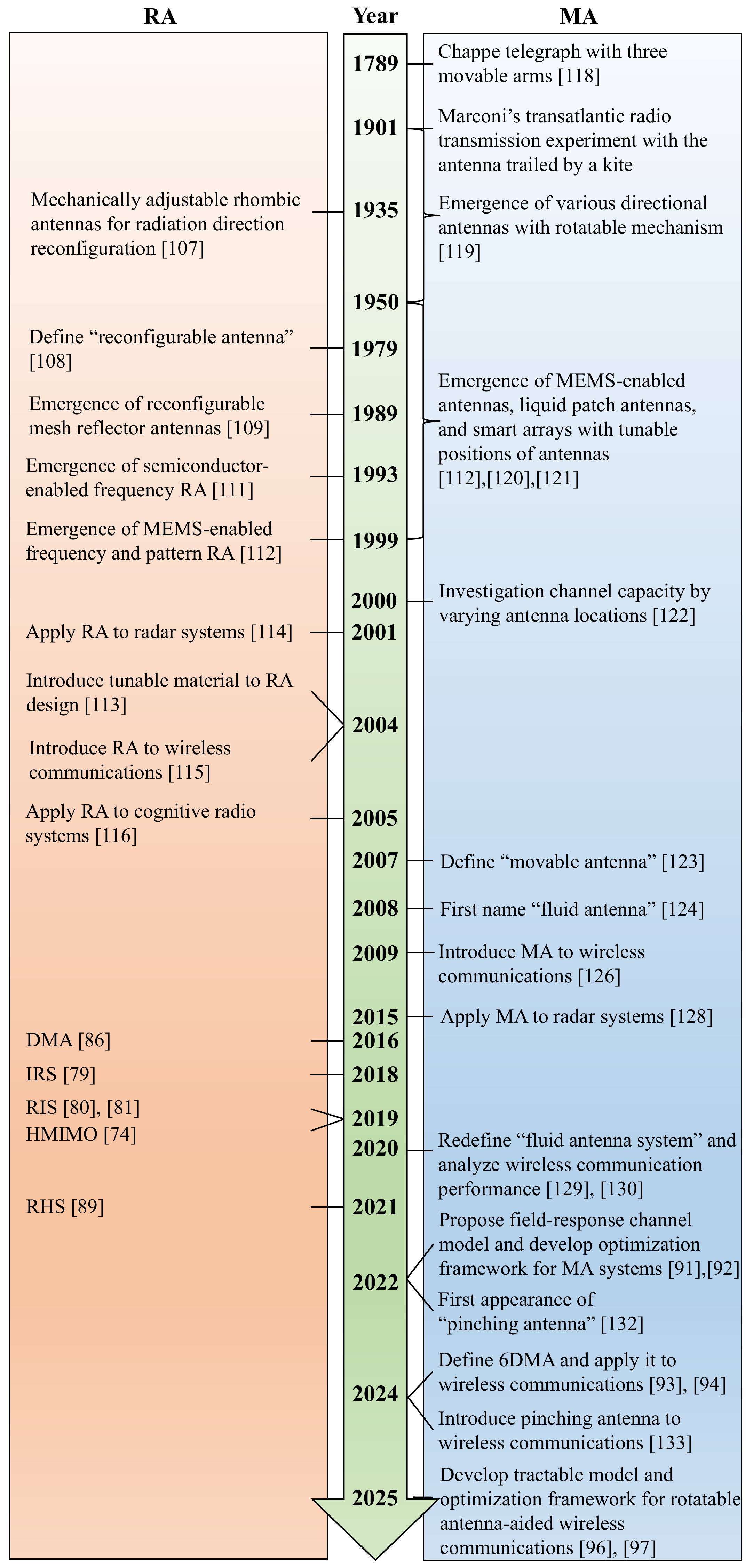}
		\caption{{Illustration of the historical development of RA and MA.}}
		\label{Fig_history}
	\end{figure}

	\begin{figure*}[!t]
		\centering
		\includegraphics[width=140mm]{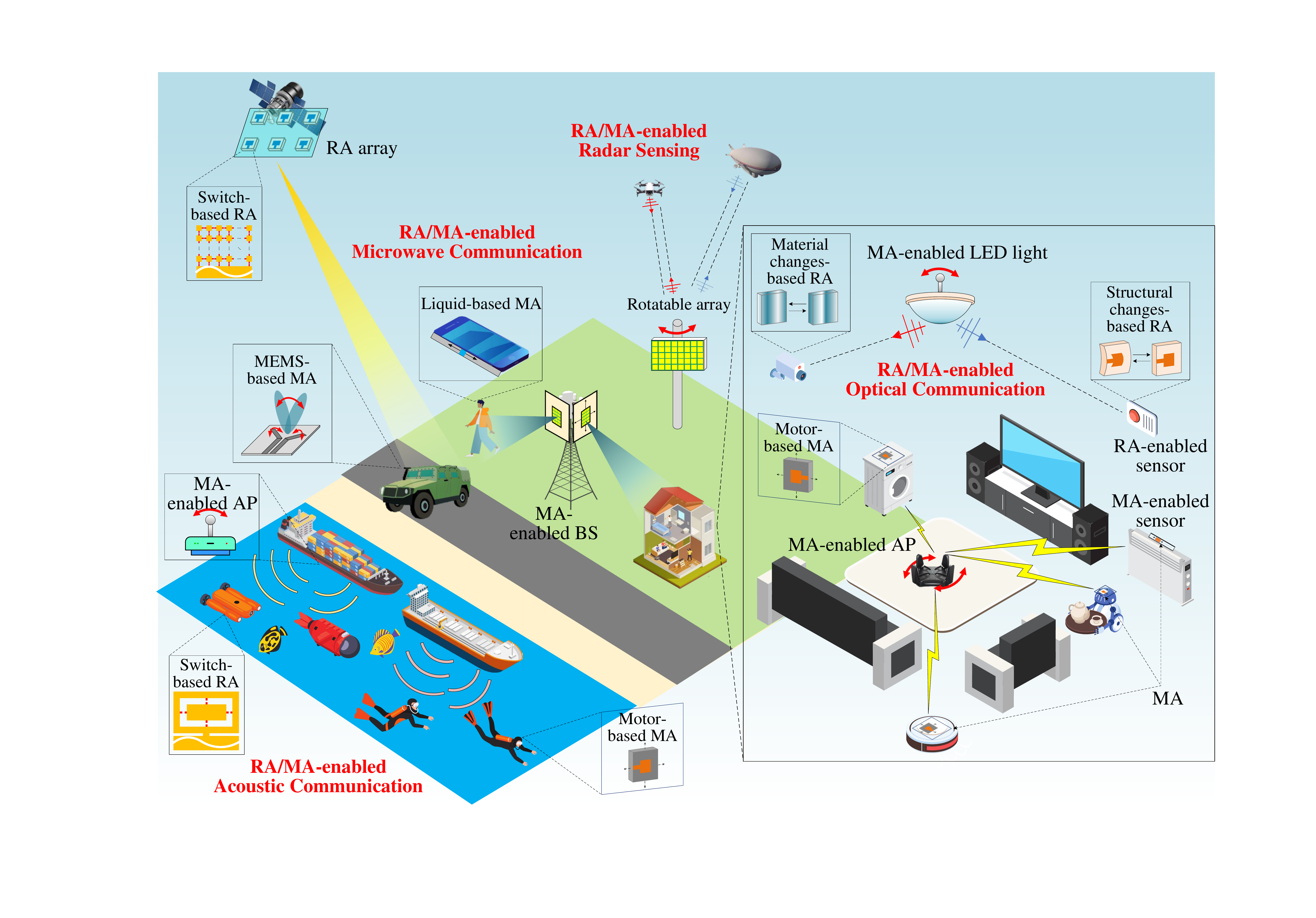}
		\caption{Application scenarios of RA/MA-aided wireless networks. RA and MA can be applied in various wireless communication and sensing scenarios, including microwave, optical, and acoustic communications, as well as radar, imaging, and wireless sensor systems. Different implementation approaches for antenna movement and reconfiguration can be adopted, depending on specific performance requirements and hardware constraints.}
		\label{Fig_Applications}
	\end{figure*}

	\subsection{Historical Development}
	\label{subsec:intro_history}
	The reconfigurable capabilities of antennas have been exploited to enhance wireless system performance almost throughout the history of communication and sensing technologies. An early precursor to modern RAs emerged in the 1930s, involving mechanically adjustable rhombic antennas, which, by modifying the wire geometry using motors and weights, effectively steered the radiation beam in elevation \cite{Bruce1935Experiments}. The term ``reconfigurable antenna'' gained formal traction in 1979, characterized by the ``ability to adjust beam shapes upon command'' and demonstrated through a multi-beam system designed to dynamically alter coverage for satellite communications \cite{Matthews1979Technology}. In 1989, the reconfigurable mesh reflector antennas were introduced for adaptive sidelobe nulling \cite{clarricoats1989reconfigurable}. By dynamically adjusting the reflector's surface profile, these systems could generate deep radiation pattern nulls to suppress co-channel interference, thereby significantly enhancing signal quality while preserving main beam gain \cite{Monk1995Adaptive}.
	
	A significant step towards electronic control occurred in the mid-1990s with the practical application of semiconductor devices such as PIN diodes and varactor diodes. These components enabled the electronic switching and tuning of antenna elements and feeding networks, paving the way for frequency-RAs. This capability was particularly valuable for frequency-agile military systems, enhancing both communication link robustness and sensing applications such as frequency-hopping radar \cite{Roscoe1993Tunable}. Further miniaturization and integration arrived in the late twentieth century, spurred by advances in MEMS \cite{chiao1999mems}. This technology inspired the development of MEMS-integrated RAs capable of tuning both radiation patterns and frequency responses with greater precision and lower loss \cite{chiao1999mems}. Alternative approaches using tunable materials also gained prominence starting in the early 2000s. Research explored materials like liquids and liquid metals whose EM properties could be altered, offering different mechanisms for achieving antenna reconfigurability \cite{Kosta2004liquid}. Since the early 2000s, frequency/radiation/polarization-RAs were explored for space-based radar \cite{Bernhard2001Stacked} as well as to enhance spatial diversity, reduce interference, or improve capacity in MIMO communication systems \cite{2004-ComMag-BA}. Around 2005, the rise of cognitive radio provided a strong impetus for RA development in communications. Cognitive radios require sensing the spectral environment and then adapting transmission parameters, including operating frequency, making frequency-RAs essential for realizing practical cognitive radio systems \cite{haykin2005cognitive}. Since the mid-2010s, advancements in metamaterials have enabled new mechanisms for achieving antenna reconfigurability, leading to the development of architectures such as HMIMO \cite{renzo2019smart}, IRS/RIS \cite{liaskos2018new, wu2019IRS, Huang2019RIS}, DMA \cite{pulido2016application}, and RHS \cite{Deng2021reconf}.
	
	In parallel, exploiting antenna movement to enhance wireless system performance has an even longer history \cite{zhu2024historical}. Early precursors include the Chappe telegraph's mechanical arms (1790s) for signaling \cite{holzmann1995early} and Marconi's kite-towed wire antenna for the first transatlantic radio reception (1901). Subsequently, directional antennas like Yagi and parabolic reflectors were often mounted on rotatable platforms for beam steering in communication and radar \cite{balanis2016antenna}. The late 20th century brought advances such as MEMS-enabled tunable antennas \cite{chiao1999mems} and optimized element positioning in arrays \cite{Lewis1983sidelobe, Ismail1991nullst}, further highlighting the benefits of adaptive configurations.
	
	A pivotal theoretical contribution emerged in 2000, which investigated the optimization of multi-antenna channel capacity by varying antenna locations \cite{chiurtu2000varying}. Building on this legacy, the term ``movable antenna'' was formally discussed in antenna monograph by 2007 \cite{balanis2007modern}. Around the same time, the term ``fluid antenna'' emerged \cite{Tam2008fluid}, initially describing antennas using liquid radiators \cite{kosta1989liquid, Kosta2004liquid} but conceptually related to MA through potential material movement. To the best of our knowledge, it was in 2009 when the first rigorous investigation of MA-aided wireless communication systems was conducted \cite{zhao2009single}, where the spatial diversity gain of a single receive MA was evaluated based on spatial-correlation channel model under rich-scattering conditions. While antenna movement has long been integral to radar techniques like synthetic aperture radar (SAR) \cite{Moreira2013SAR}, with MA prototypes showing radar imaging benefits \cite{Zhuravlev2015experi}, its dedicated application in wireless communications saw less focus until recently.
	
	In 2020, the authors in \cite{wong2020limit, wong2020fluid} expanded the definition of fluid antenna systems (FAS) to encompass any position- or shape-flexible antenna  and analyzed the spatial diversity gain of a single receive fluid antenna using a simplified form of the spatial correlation channel model in \cite{zhao2009single} under the assumption of rich scattering. Following its original definition, the MA-aided wireless communication system with joint transmit and receive antenna 2D movement was investigated in 2022  \cite{zhu2022MAmodel, ma2022MAmimo}, which was then extended to 3D antenna movement in \cite{zhu2023MAMag,zhu2023MAmultiuser}. By introducing the new antenna rotation DoF, 6DMA system was defined in 2024 to achieve the highest flexibility  in antenna movement through integrating 3D position and 3D orientation/rotation adjustments \cite{shao20246DMA, shao2024Mag6DMA, shao2024discrete}, thus achieving further significant capacity enhancement without increasing or even reducing the number of antennas. In late 2024 and 2025, rotatable antenna technology \cite{zheng2025rotatable,zhengtian2025rotatable} emerged as a new variant that focuses solely on antenna boresight rotation flexibility (with fixed antenna position), offering a cost-effective and compact solution regarded as a simplified yet promising subset of MA/6DMA. Moreover, the pinching antenna, introduced in 2022, enables antenna movement or positioning by allowing radio waves to be radiated from arbitrary points along a dielectric waveguide \cite{suzuki2022pinching}. Since 2024, the pinching antenna has been applied to various communication scenarios, including single-input single-output (SISO), non-orthogonal multiple access (NOMA), and multiple-input single-output (MISO) communication systems \cite{ding2024flexible,liu2025pinching,yang2025pinching,liu2025pinchingTut,ouyang2025array}. Despite distinct origins and different implementation methods, MA, FAS, and pinching antenna share fundamentally similar principles of flexible antenna positioning, and they have attracted growing interest in the communication community in recent years \cite{zhu2024historical}. The historical development of RA and MA technologies is summarized in Fig.~\ref{Fig_history}.

	\subsection{Applications}
	\label{subsec:intro_apps}
	By leveraging reconfigurable radiation characteristics via RAs and adaptive antenna movement via MAs, these antenna technologies offer significant opportunities to enhance the antenna flexibility, performance, and efficiency across various communication and sensing applications. The ability of RA or MA to dynamically alter antenna's internal or external properties unlocks new DoFs compared to conventional antenna systems. As shown in Fig.~\ref{Fig_Applications}, RA and MA can be applied jointly or individually to improve system capabilities in various wireless communication and sensing scenarios.
	
	\subsubsection{Microwave Communications}
	Conventional microwave communication systems often employ fixed and non-reconfigurable antennas, limiting their ability to adapt to dynamic channel conditions caused by mobility, obstructions, and/or environmental changes. This can lead to suboptimal link quality and inefficient spectrum utilization. In contrast, RA systems can electronically steer beams or change polarization to optimize links, while MA systems add the capability to physically reposition or reorient antennas. The adaptability of RA and MA allows for precise beam alignment, enhanced channel gain, and effective interference mitigation in microwave links, improving throughput and reliability even in challenging propagation environments \cite{ma2022MAmimo, shao20246DMA, shao2024discrete, shao2024Mag6DMA, xiao2023multiuser, wu2023movable, Yang2024movable, zhu2023MAmultiuser, Hu20242024twotimeMA, zheng2024twotimeMA}. Moreover, for low-earth orbit (LEO) satellites, implementing reconfigurable array geometries at the satellites or ground stations facilitates more adaptable beamforming, leading to enhanced coverage and interference suppression for terrestrial users \cite{ZhuLP_satellite_MA, lin2024power, wang2025joint}. For unmanned aerial vehicles (UAVs), the platform's inherent large-scale mobility can be leveraged alongside MAs to ensure robust, uninterrupted 3D connectivity between aerial and ground-based communication systems \cite{kuang2024movableISAC, ren20246DMAUAV, liu2024uav6DMA}.
	
	\subsubsection{Optical Communications}
	Free-space optical (FSO) communication relies on highly directional beams, making beam alignment critical and susceptible to atmospheric turbulence, building sway, and/or platform vibrations. Traditional FSO systems often use complex and slow mechanical gimbals for alignment. RA techniques can offer faster beam steering capabilities, while MA can precisely adjust the antenna (optical transceiver) position and/or orientation to maintain optimal alignment and maximize received power \cite{zhu2023MAMag,WangHH_interference_MA}. Using RA for beam steering and using MA for position and/or orientation alignment can significantly improve link stability, reduce pointing errors, and enhance the throughput of FSO systems, especially for mobile platforms or long-distance links.
	
	\subsubsection{Acoustic Communications}
	Underwater or through-medium acoustic communication faces challenges like severe multipath propagation, limited bandwidth, and slow sound speed, making reliable communication difficult with conventional antenna systems. RA could enable tunable beam patterns or operating frequencies, while MA can exploit spatial diversity and find locations with favorable channel conditions, mitigating multipath fading and improving the signal-to-noise ratio (SNR) \cite{Wang2018maritime, Alqurashi2023maritime}. The adoption of RA or MA allows acoustic communication systems to adaptively optimize transmission and reception, enhancing data rates, reliability, and range in complex underwater or reverberant environments.
	
	\subsubsection{Radar}
	Radar systems with conventional non-reconfigurable/movable antennas/arrays have limitations in terms of spatial resolution and interference mitigation, typically requiring a large number of antennas to achieve high performance. RA can provide adaptable beam shapes, while MA can reconfigure the array geometry dynamically. By optimizing antenna positions, MA arrays can synthesize a larger virtual aperture than the physical array size, significantly enhancing angular resolution and target detection capabilities without increasing the antenna number \cite{ma2024MAsensing, Chen2024moving, shao2024exploiting}. RA/MA radar systems can adapt their configuration for different tasks, suppress clutter and interference more effectively, and achieve higher resolution compared to conventional antenna-based radars. Moreover, MAs are a key enabling technology for ISAC, improving both sensing accuracy and communication performance. By reconfiguring antenna positions, ISAC systems can flexibly manage the trade-off between communication quality and sensing precision, thereby adapting to varying operational requirements \cite{kuang2024movableISAC, ma2025MAISAC, li2024MAISACMag, WuHS_MA_RIS_ISAC, lyu2024flexibleISAC, peng2024jointISAC, wang2024multiuser, guo2024movable, MaY2024movableISAC, zhou2024fluidISAC}.

	\begin{figure*}[t]
		\begin{centering}
			\includegraphics[width=16cm]{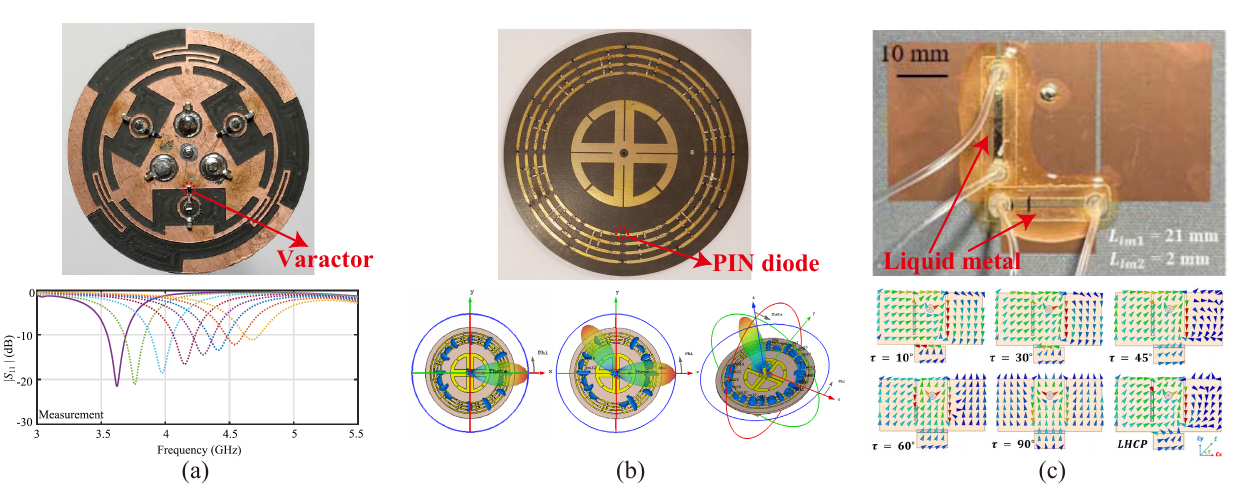}
			\par\end{centering}
		\caption{\label{fig:Classification-and-architectures-of-RA}Classification
			and architectures of RAs. (a) Frequency-RA with continuous frequency tuning capability \cite{2022-TAP-FeRA};
			(b) Pattern-RA with 3D beamforming capability
			\cite{HKUST_360}; (c) Polarization-RA
			with multi-polarization generation capability \cite{2020-TAP-contiounespol}.}
	\end{figure*}
	
	\subsubsection{Imaging}
	Microwave or mmWave imaging systems often suffer trade-offs among resolution, field-of-view, and system complexity when using conventional antenna arrays. RA allows for scanning beams or adapting patterns electronically. MA enables the physical reconfiguration of the imaging array, potentially creating sparse array configurations with optimized antenna positions for specific imaging tasks \cite{venkatesh2022origami}. This can achieve high resolution over a wide area with fewer antennas than a dense array with fixed-position antennas (FPAs), reducing system cost and complexity. By dynamically adjusting antenna positions for MA-enabled systems or radiation patterns for RA-enabled systems, imaging systems can adapt to different scenes, improve image quality, suppress artifacts, and offer enhanced capabilities for applications such as security screening, medical imaging, or non-destructive testing. This is conceptually similar to SAR \cite{Moreira2013SAR} while allowing for more flexible and real-time aperture synthesis.
	
	\subsubsection{Sensors}
	Wireless sensor networks (WSNs) often deploy sensors with FPAs, which may not be optimally positioned for communication links or sensing tasks, especially in dynamic environments. RA can allow sensor nodes to adapt their communication patterns, while MA permits physical repositioning to improve link quality to data sinks or optimize sensing coverage for specific phenomena \cite{liu2024four,raza2024precision}. For distributed sensing tasks, MA-enabled sensors could coordinate their positions to form optimal geometries for source localization or environmental field mapping. The adaptability provided by RA and MA enhances the resilience, efficiency, and sensing accuracy of WSNs, particularly for mobile sensors or applications requiring targeted monitoring in complex terrains.

	\section{Antenna Architectures}
	\label{sec:architectures}
	\subsection{RA Architectures}
	\subsubsection{Reconfigurable Antenna-Element}
	An RA is well known as a single antenna element
	capable of dynamically changing its operating parameters, such as
	radiation pattern, operating frequency band, and polarization, to
	better adapt to different wireless channels \cite{Christodoulou2012reconfig,2007-Jennifer-RAbook,costantine2015reconfigurable}.
	Compared to traditional phased arrays, which require multiple expensive
	RF chains and phase shifters, RAs offer a low-cost and compact alternative.
	Moreover, unlike conventional antenna elements with fixed characteristics,
	RA provides more design freedom and diversity to wireless systems.
	These advantages have led to growing interest in the application of
	RAs in modern wireless systems.
	
	Designing an RA is not just about making a single antenna. It involves
	creating multiple functions/antennas and combining them
	in one structure. In addition, designing RAs involves applying knowledge
	from antenna theory, RF and direct current (DC) circuits, and tunable materials. These
	make RA design a wide and promising research area.	
	In this subsection, we review the current progress of RAs, with a focus
	on their classification, reconfiguration methods, and key design challenges.
	
		\begin{table*}[t]
		\caption{\label{tab:Key-Characteristics-of-RA}Key Characteristics of Representative
			Reconfigurable Devices and Materials Used in RAs.}
		
		\centering{}%
		\scalebox{0.9}{
			\begin{tabular}{|c|c|c|c|c|c|}
				\hline 
				Reconfigurable Methods & Tuning Type & Tuning Speed & Tuning Voltage & Power Consumption & Applied Frequency\tabularnewline
				\hline  
				PIN Diode & Discrete & Fast (ns) & Low & Moderate & Low to moderate\tabularnewline
				\hline 
				Varactor & Continuous & Fast (ns) & High & Low & Low\tabularnewline
				\hline 
				MEMS & Discrete/Continuous & Moderate (µs$\sim$ms) & High & Low & Low to high\tabularnewline
				\hline 
				Liquid Metal & Continuous & Slow (ms$\sim$s) & High & High & Low to high\tabularnewline
				\hline 
				Liquid Crystal & Continuous & Slow (ms$\sim$s) & Low & Low & High\tabularnewline
				\hline 
				Mechanical Movement & Discrete/Continuous & Slow (ms$\sim$s) & High & High & Low to high\tabularnewline
				\hline 
		\end{tabular}}
	\end{table*}
	
	\paragraph{Classification}
	
	In terms of the reconfigurable parameters, RA can be classified
	into pattern-RA, frequency-RA,
	and polarization-RA. Some representative RA architectures
	and performance are shown in Fig. \ref{fig:Classification-and-architectures-of-RA}. 
	
	\textit{Pattern-RA}: Representative methods for
	realizing pattern-RAs include pixelated surfaces
	\cite{HKUST_360,2024-TAP-PRA,2015-TAP-PRApixel,jiang2021pixel,RPA_BF,2017-TAP-PArisa,chen2025remaa},
	electronically steerable parasitic array radiators (ESPAR) \cite{2022-TWC-Han,2012-TAP-ESPAR-Gao,1978-Harrington,HKUST_ESPAR,2004-TAP-espar_SINR},
	reconfigurable metasurfaces/metamaterials/frequency selective surface (FSS) \cite{2013-FSSs-dualbeam,2022-TAP-VP,2013-TAP-FSS_10beam6gain,2017-TAP-FSSs_patch9dBi,2022-AWPL-Dual-pol-meta},
	and selected driven antennas \cite{2009-TAP-6espar,2018-AWPL-ARC_dipole4beam,2017-AWPL-pixel4beam,2018-TAP-MEdiople4beam}. { Although pixelated surfaces and reconfigurable metasurfaces/metamaterials share similar principles, they are distinguished in this paper based on the activeness of the radiating elements and the adopted design methodology. A pixelated surface specifically refers to a structure in which the radiating aperture is discretized into small metallic elements. A pixelated surface realizes its functionality by electrically connecting or disconnecting these discrete metallic pixels, thereby modifying the antenna topology and the associated surface current distributions. In contrast, reconfigurable metasurfaces and metamaterials achieve reconfiguration by tuning the EM properties of individual subwavelength unit cells, rather than by electrically connecting different cells. This enables controlled manipulation of wave propagation and scattering through engineered effective material responses.}
	The key design strategy lies in manipulating the current distribution
	on the antenna's radiating aperture. Currently, most single-port pattern-reconfigurable
	antennas can only provide a limited number of reconfigurable patterns
	in a 1D elevation or azimuth plane. For future designs, enhanced
	beamforming capabilities, such as 2D or even 3D beam-steering or beamforming,
	are desirable \cite{HKUST_360}. Highly pattern-reconfigurable
	antenna offers increased design freedom for wireless systems. Additionally,
	developing compact multi-port pattern-RAs with
	independent beam control at each port is highly promising and can
	directly enhance channel capacity in wireless communications \cite{2024-TAP-PRA,2024-Eucap-Dirk,2023-OJAP-multi-portPRA}.
	
	\textit{Frequency-RA}: The most common approach
	for designing frequency-RAs is to use tunable materials
	for adjusting the effective antenna dimensions (i.e., resonant current
	path), thereby shifting the resonant frequency \cite{2018-TAP-FRE,2008-AWPL-FPRA,2009-AWPL-FrePRA,2022-TAP-FeRA,2006-TAP-FRe}.
	This can be achieved by extending the antenna length with liquid metal,
	altering the substrate\textquoteright s dielectric constant using
	liquid crystal, or loading parasitic elements with PIN diodes or varactors.
	These methods typically allow up to 50\% frequency tuning. To surpass
	this limit, designs that switch between distinct radiating modes with
	widely separated resonant frequencies can be employed. A recent example
	achieves a 83\% tuning range by switching between planar inverted-F antenna (PIFA) and patch modes,
	as reported in \cite{2021-TAP-FRA} and \cite{2024-AWPL-83tunning}.
	
	\textit{Polarization-RA}: The key for designing
	polarization-RAs is exciting orthogonal modes
	and controlling their phase differences. This can be achieved in various
	ways, for example, using etched slots on the radiating patch to convert
	linear polarization (LP) to CP \cite{2020-TAP-pol.GuoYJ},
	employing reconfigurable defected ground structures (DGS) to switch
	between right-hand CP (RHCP) and left-hand CP (LHCP) \cite{2020-OJAP-pol}, or adopting rotationally
	symmetric geometries to realize multiple linear polarizations \cite{2023-TAP-16pol}.
	Continuous LP reconfiguration is also possible by tuning the relative
	amplitude of the orthogonal modes with tunable materials \cite{2020-TAP-contiounespol}.
	
	Since the three RA parameters are independent,
	they can be jointly designed to create multi-functional antennas \cite{2015-TAP-PolPRA,2020-TAP-polPRA,2004-ComMag-BA,2006-TAP-patter-Fre,2014-TAP-FPRA,2014-TAP-metasurfaceFPRA,MRA_2}. This represents a more advanced form where multiple antenna properties (e.g., frequency and pattern, or pattern and polarization) are adjusted simultaneously or sequentially \cite{MRA_1, MRA_2}. It provides enhanced flexibility by allowing the antenna to adapt to complex operating scenarios requiring optimization across several domains. Compound reconfiguration enables synergistic benefits, such as concurrently tuning to the best frequency band while steering the beam towards the intended user for optimal link quality.

	\paragraph{Reconfigurable Methods}
	
	Beyond antenna structure design, the choice of reconfiguration method
	is also critical. In many cases, the reconfigurable material determines
	the performance limits of the antenna. In the following, we review some of the
	most commonly used reconfigurable devices and materials.
	
	\textit{Semiconductor Diode}: PIN diodes and varactor diodes are the
	most commonly used devices in RAs. They can change
	the antenna\textquoteright s performance by providing different impedance
	values. A PIN diode acts like an RF switch. Its ON/OFF states work
	like a short or open circuit. PIN diode responds very fast with several
	nanoseconds switching speed. It only needs a small DC voltage to operate
	and is very compatible with digital controllers like field-programmable gate arrays (FPGAs) \cite{2020-APS-mmW}.
	Varactor diodes offer continuous tuning by changing their reactance,
	which allows more precise control. For antenna design, a wide reactance
	tuning range in varactors is especially useful \cite{2023-OJAP-varactor}.
	Driven by the laser, photodiode can also be utilized when the
	diode numbers are externally large. This could help remove the massive
	DC bias lines placed near the antenna geometry \cite{2004-TAP-optial-controlled}.
	
	\textit{Liquid Metal}: Liquid metal, such as mercury (Hg), is a fluidic
	conductive material whose position can be controlled using microfluidic
	channels and micropumps \cite{2020-Liqiud_metal,2018-AWPL-liquid_metal,2012-TAP-liquidmetalBApixelgroup,2024-Wangyi-liquidmetal}.
	One major advantage is its high conductivity, which results in very
	low loss. It can act directly as the antenna radiator and offer great
	tuning flexibility. {Liquid metal performs well in higher frequency bands due to its high electrical conductivity and the resulting lower insertion loss compared to semiconductor switches in mmWave and THz frequency bands.}
	However, the main challenges of liquid metal are the complexity of
	the supporting systems, such as the microfluidic channels and pumps,
	which are difficult to integrate into antenna designs. Additionally,
	the slow movement speed of the fluid may limit the performance of
	wireless communication systems.
	
	\textit{Liquid Crystal}: Liquid crystal is a special material whose
	effective permittivity can be tuned by rotating its rod-like molecular
	structure. It is commonly used as a continuously tunable phase shifter
	in RA or reflect-array designs \cite{2015-TAP-LCPRS,2022-AWPL-LC,2014-TAP-LCpatch}.
	Due to its limited permittivity tuning range and thin thickness, it
	is typically used at very high frequencies (e.g., mmWave/THz) \cite{2015-TAP-LC}.
	
	\textit{Mechanical Movement}: RF MEMS
	switches are among the most widely used technologies in RA design \cite{2012-TAP-frankieMEMS,2006-TAP-MEMSJenifer,2006-TAP-MEMSCrhis}.
	Compared to other switching devices such as PIN diodes and field-effect
	transistors (FETs), MEMS switches offer superior performance in terms
	of insertion loss, isolation, and power consumption. Other mechanical
	methods include reconfigurable aperture antennas that use actuators
	to physically adjust the antenna structure \cite{2021-AWPL-mechinal}
	and \cite{2022-TAP-mechanical}. {It is worth noting that although MEMS-based RAs typically require relatively high tuning voltages, the corresponding tuning current is almost zero. As a result, their power consumption remains very low, which is fundamentally different from mechanical movement-based reconfiguration methods, where the power consumption remains relatively high during mechanical movement.}
	
	There are also many other reconfigurable methods that can be utilized in RA design, such as origami \cite{2021-TAP-origami},
	phase change materials (PCM) \cite{2020-OJAP-PCM}, and shape memory
	alloy \cite{2024-shapemeomery-proceeding}, etc. A performance summary table is given
	in Table \ref{tab:Key-Characteristics-of-RA}.

	\begin{figure*}[!t]
		\centering
		\includegraphics[width=160mm]{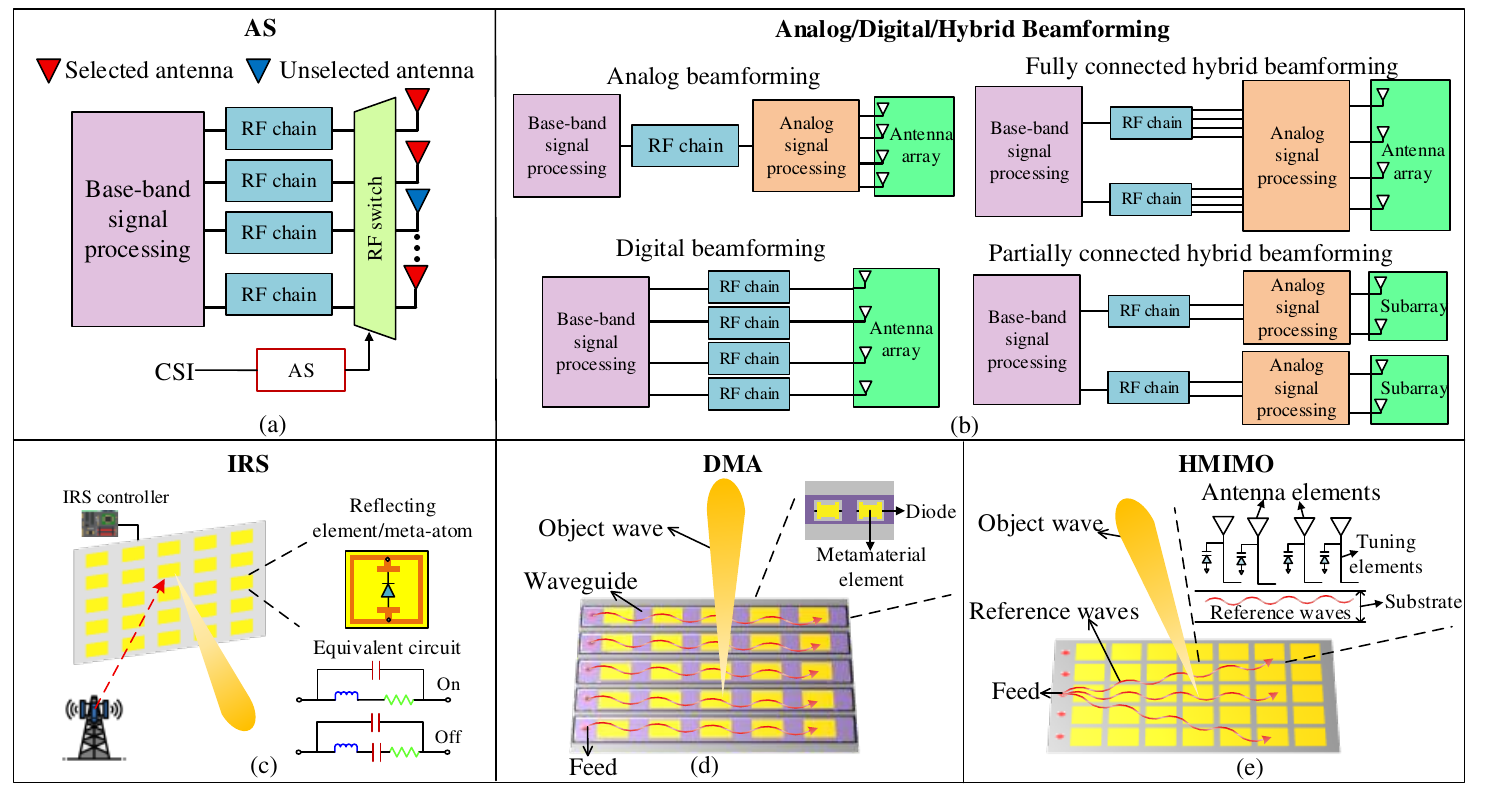}
		\caption{Illustration of typical reconfigurable methods for RA arrays.}
		\label{Fig_RA_array}
	\end{figure*}

	\paragraph{Design Challenges and Possible Solutions}
	
	Despite years of research on RA, many fundamental scientific problems
	remain open and continue to attract academic interest. In the following, we highlight
	some key design challenges and potential solutions.
	
	\textit{New RA Geometry Design}: Most previous RA designs are single-port.
	Multi-port multi-mode RA designs are highly promising for wireless
	systems, as increasing the number of ports within a given area can
	directly improve data rates. However, the design process of multi-port
	multi-mode RAs is highly challenging. It requires independent control
	of the antenna parameters at each port while maintaining a high level
	of port isolation. A promising approach to address this challenge
	is to deeply integrate microwave theories, such as multi-port theory,
	RF circuit theory, and characteristic mode theory, into the antenna
	design process \cite{2024-TAP-PRA}, \cite{2024-Eucap-Dirk}, \cite{2023-OJAP-multi-portPRA}.
	
	\textit{Reconfigurable Material}: Reconfigurable materials are key
	components in RA design. Currently, PIN diodes and varactors are the
	most commonly used tunable materials in RAs. However, they face several
	challenges, including the accuracy of equivalent parameter modeling,
	performance stability across wide operating bands, and performance
	degradation at high frequencies such as mmWave and THz
	bands. Other tunable materials, such as liquid metal, liquid crystal,
	and MEMS switches, offer better frequency stability and can operate
	effectively in high-frequency bands. However, their switching
	speed remains a limitation, making it difficult to adapt to rapidly
	changing wireless channels. In summary, there is a need to develop
	advanced tunable materials that combine fast switching speed, a wide
	impedance tuning range, stable frequency response, and high performance
	at ultra-high frequencies \cite{2021-AWPL-mmWPRA,2020-NE-THz-diode,2023-APS-mmwswitch}.
	
	\textit{Modeling RA Parameters into Wireless Systems}:
	Although the EM performance of RAs has been widely studied,
	few works have focused on directly modeling RA parameters into wireless
	system performance. Therefore, it is promising for future research
	to integrate EM theory with wireless communication theory.
	System level optimization, such as capacity maximization based directly
	on RA parameters, holds great potential \cite{2022-TWC-Han,HKUST_AP,2022-VTC-reconfigruable-MIMO,2021-TVT-Pixel}.
	This approach can help bridge the gap between these two research domains.
	
	\subsubsection{Reconfigurable Antenna-Array}
	In this subsection, we review recent developments in RA arrays, with particular attention to their classification, implementation strategies, and associated design challenges.
	
	\paragraph{Classification}
	From a structural perspective, RA arrays can be divided into two main configurations~\cite{castellanos2025embracing,10077503,HKUST_360}.
	One configuration consists of numerous independently RA elements, each capable of adjusting parameters such as frequency, radiation pattern, or polarization. By jointly controlling all the elements, fine-grained control of the overall array can be achieved. However, this configuration typically incurs greater control complexity and higher power consumption. Alternatively, the array can be divided into multiple subarrays, where groups of elements share common hardware resources (e.g., RF chains and phase shifters), thereby simplifying system design and reducing hardware complexity and power consumption, albeit at the cost of reduced fine-grained reconfigurability and control accuracy. Accordingly, an appropriate trade-off or hybrid configuration can be flexibly adopted depending on system performance requirements, as well as power and complexity constraints.

	\paragraph{Reconfigurable Methods}
	Various implementation methods can be employed to unlock the potential of RA arrays. As illustrated in Fig.~\ref{Fig_RA_array}, those typical techniques include:
	
	\textit{AS}: This is considered as a low-complexity reconfiguration method where only a subset of antennas is activated based on channel conditions, thus reducing the number of RF chains and power consumption~\cite{Sanayei2004antennas,8234671,6725592,7574379}. By selecting the ``best'' antennas, AS can retain much of the performance gain of the large array while simplifying the hardware implementation. Notably, the AS technique can be regarded as a form of spatial reconfiguration, which selectively activates antenna elements or subarrays based on their positions within the array. It is especially useful in mMIMO systems with limited RF resources, 
	aiming to achieve a better balance between throughput and cost. However, AS requires fast and adaptive switching algorithms to cope with channel variations, and it offers less beamforming gain than the full-array control. Despite these limitations, AS remains an efficient solution in power-limited or hardware-constrained scenarios.
	
	\textit{Analog/Digital/Hybrid Beamforming}: Analog, digital, and hybrid beamforming schemes are key techniques employed in RA arrays to steer and reshape radiation patterns, each characterized by distinct trade-offs among performance, flexibility, and implementation complexity~\cite{5447703}. Analog beamforming utilizes phase shifters and variable gain amplifiers (VGAs) to manipulate the phase and amplitude of signals directly in the RF domain, providing a low-complexity and energy-efficient solution for beamforming. However, analog beamforming typically generates a single beam that serves a specific user or spatial region. As a result, it inherently lacks the ability to independently form multiple beams, which is essential for serving multiple users or supporting spatial multiplexing in MU-MIMO systems. Therefore, analog methods are often complemented by digital beamforming, which operates at the baseband level with advanced signal processing algorithms and high-resolution converters. Digital beamforming enables precise control over beam patterns, supports multi-stream transmission, and provides robust interference mitigation, albeit at the expense of increased hardware complexity and power consumption~\cite{8353836,8883297}. Hybrid beamforming strikes a balance between the two approaches by combining analog front-end processing with a reduced number of digital processing chains. This architecture significantly lowers implementation cost and energy requirements while retaining sufficient beamforming capability, making it especially suitable for high-frequency applications such as mmWave communications~\cite{7389996,8616797}.
	
	\textit{IRS}: IRS represents a class of passive RA structures that extend traditional antenna array functionality into the wireless propagation environment~\cite{zheng2022survey,liu2021RIS,wu2024ISfor6G,10133841,9805460}. Unlike conventional arrays that actively transmit or receive signals, IRS consists of a large number of low-cost passive reflecting elements capable of dynamically tuning the phase of incident signals, collaboratively reshaping the wireless channels. This allows IRS to support key communication functions such as signal enhancement, interference mitigation, and spatial multiplexing, without relying on active RF chains or power-hungry signal processing units~\cite{zheng2020IRSOFDM,9362274}. As such, IRS bridges the gap between transceiver-side array processing and environment-side reconfiguration, providing a cost-effective and energy-efficient solution for future wireless networks such as 6G. However, integrating IRS into practical systems poses significant challenges, particularly in accurate channel estimation and passive beamforming under hardware constraints, thereby motivating extensive research on robust signal processing and system design tailored to this new class of RA arrays.
	
	\textit{DMA}: As an active RA technology, DMA has emerged as a promising hardware solution for realizing mMIMO transceivers in 6G wireless systems~\cite{Shlezinger2021DMAs}. Unlike conventional antenna arrays that rely on complex RF chains and phase shifters, DMA utilizes waveguide-fed arrays of tunable metamaterial elements to achieve beamforming and signal processing~\cite{boyarsky2021electronically}. This unique structure enables DMA to support a large number of antenna elements with significantly fewer RF chains, reducing hardware cost, power consumption, and system complexity~\cite{10938032}. Moreover, compared with conventional antenna arrays based on hybrid beamforming, DMA can not only accommodate more antenna elements within the same aperture but also eliminate the need for complex corporate feeds and/or active phase shifters. In addition, extensive research has demonstrated that DMA can approach the performance of fully-digital mMIMO systems, even under limited RF chain constraints~\cite{9272351,Shlezinger2019DMA,9762020}. Despite their potential, key challenges remain in areas such as efficient channel estimation and hardware implementation at mmWave and THz bands. Addressing these challenges will be essential to fully realize the advantages of DMA in practical deployments.
	
	\textit{HMIMO}: As a forward-looking extension of RA array technology, HMIMO represents a transformative approach for future wireless networks by redefining how EM waves are generated, manipulated, and received~\cite{Huang2020hologr}. Unlike conventional mMIMO relying on discrete antenna arrays with half-wavelength spacing, HMIMO utilizes nearly continuous, ultra-thin surfaces composed of densely packed, sub-wavelength elements or metasurfaces~\cite{gong2023holographic}. These surfaces, either active or passive operation, enable high-resolution beamforming, wave focusing, and polarization control directly in the EM domain. In particular, active HMIMO integrates RF chains and signal processing for direct EM-domain beamforming, while passive HMIMO (often realized as IRS~\cite{zheng2022survey}) manipulates waves with low power consumption, suitable for cost-effective deployment~\cite{Pizzo2020spatial}. Therefore, HMIMO transforms the wireless medium from a passive channel into a programmable environment, offering substantial gains in signal quality, energy efficiency, and capacity. { It is worth emphasizing that IRS can be considered as a passive and reflective RA array technology that does not actively generate RF signals, but instead reshapes the wireless propagation environment by controlling the reflection of incident waves. This is fundamentally different from active and radiative architectures, such as DMA and active HMIMO, which directly transmit and process RF signals.} From another perspective, HMIMO also bridges Shannon’s information theory and Maxwell's equations, thereby fostering the development of EM information theory~\cite{10041914}. Despite its promise, HMIMO faces challenges in hardware scalability, mutual coupling, near-field channel modeling, and EM-aware signal processing. Nonetheless, its ability to unify communication, sensing, and control in a reconfigurable EM environment positions HMIMO as a foundational technology for future intelligent and immersive systems.
	
	\begin{figure*}[!t]
		\centering
		\includegraphics[width=150mm]{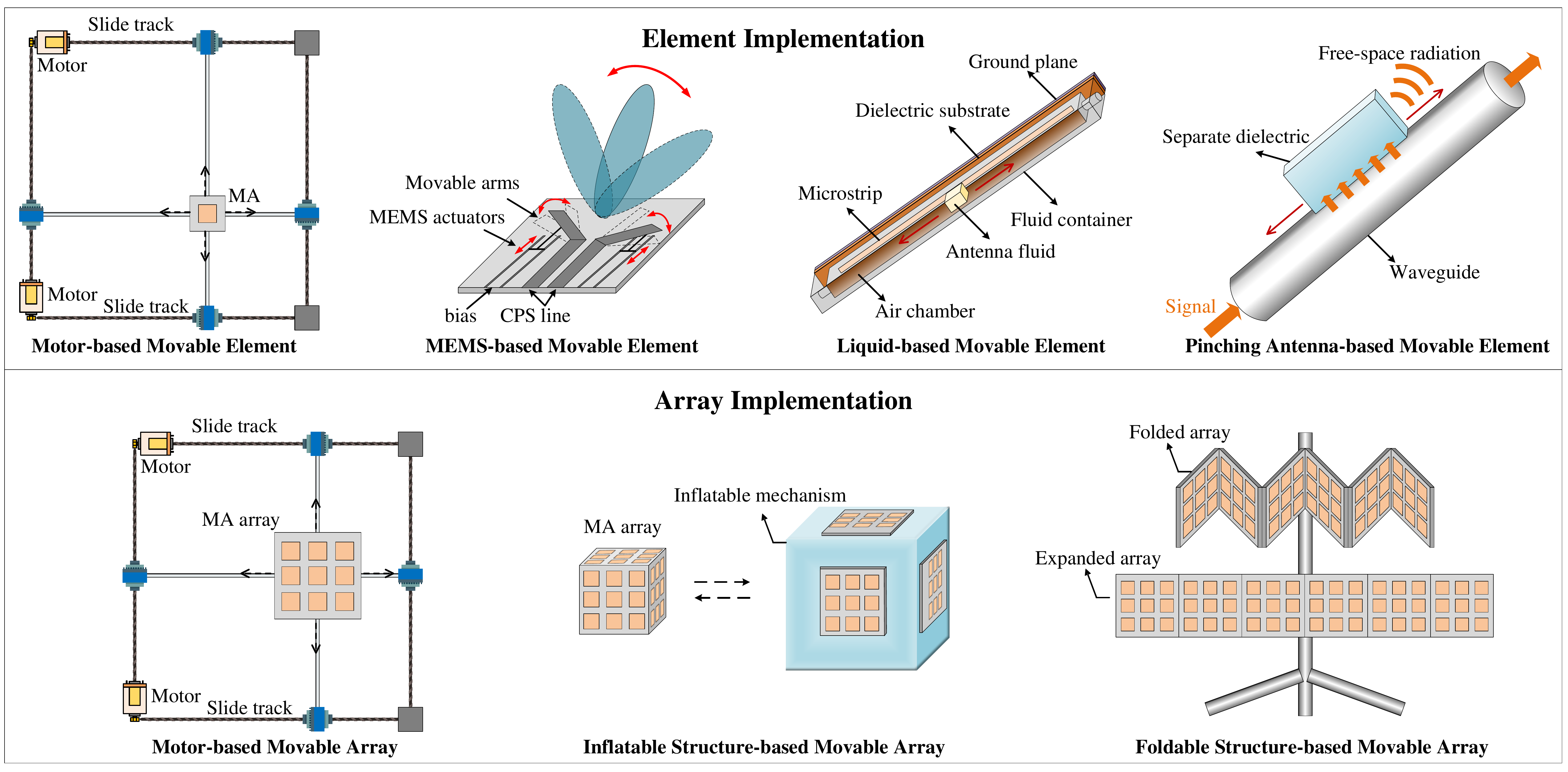}
		\caption{Illustration of typical implementation methods for realizing antenna movement and reconfiguration.}
		\label{Fig_Implementations}
	\end{figure*}
	
	\paragraph{Design Challenges and Possible Solutions}
	From a structural perspective, a key design challenge for RA arrays lies in managing the trade-off between control granularity and system complexity. While employing numerous independently reconfigurable elements offers the finest control over the array's radiation properties, this approach often leads to prohibitive control complexity and high power consumption \cite{castellanos2025embracing,10077503,HKUST_360}. A possible solution is to partition the array into subarrays that share common hardware resources, which simplifies the design and reduces power, albeit at the cost of less precise reconfigurability. Furthermore, various implementation strategies present their own challenges. For instance, AS is a low-complexity method but requires fast, adaptive algorithms and provides less beamforming gain than full-array control \cite{Sanayei2004antennas,8234671,6725592,7574379}. Hybrid beamforming strikes a balance between performance and cost but requires careful partitioning of analog and digital resources \cite{7389996,8616797}. More advanced architectures like IRS and DMA face significant hurdles in efficient and accurate channel estimation \cite{zheng2022survey,liu2021RIS,wu2024ISfor6G,10133841,Shlezinger2021DMAs}, while holographic MIMO introduces challenges in hardware scalability, mutual coupling, and near-field modeling \cite{Huang2020hologr}.
	
	In summary, different forms of RA techniques exhibit distinct functionalities and entail varying trade-offs in performance, system complexity, and implementation cost. 
	As a result, future research and deployments will benefit from integrated strategies that harness the complementary strengths of these technologies, thereby enhancing system performance and enabling flexible, efficient, and highly adaptive wireless networks.

	\subsection{MA Architectures}
	The practical realization of MA/6DMA systems is fundamentally dependent on their hardware architectures. These architectures define the methods and constraints associated with physically altering an antenna's position or orientation to adapt to the wireless environment. As illustrated in Fig.~\ref{Fig_Implementations}, the approaches to implement MA/6DMA can be broadly categorized based on whether the movement pertains to individual antenna elements or to entire antenna arrays/subarrays. Each category encompasses distinct mechanical strategies to achieve the desired spatial adaptability.
	
	\subsubsection{Movable Antenna-Element}
	Architectures for movable antenna-elements in 6DMA systems are designed to enable each individual antenna-element to change its physical position and/or orientation. This element-level control allows for precise adjustments to improve wireless communication or sensing performance. In this subsection, we review the current progress of movable antenna-element, with a focus on their classification, implementation methods, and key design challenges.

	\paragraph{Classification}
	
	Based on the available hardware-enabled DoFs, the movement of individual antenna elements in 6DMA systems is classified into two fundamental types: translation and orientation \cite{shao2025tutorial}.
	
	\textit{Translation}: Position reconfiguration involves the physical translation of an antenna element. This can be along a 1D line segment or within a given 2D/3D space \cite{zhu2023MAMag,zhu2022MAmodel}. Such movement allows the antenna's phase center to be dynamically relocated to positions in space that offer more favorable channel conditions, such as increasing signal strength or decreasing interference \cite{ding2024MAnearISAC}. 
	
	\textit{Orientation}: In addition to translation, orientation reconfiguration allows an antenna element to be rotated \cite{shao20246DMA, shao2024Mag6DMA, shao2024discrete, shao2025tutorial}. This can involve rotation about a single axis, or more complex rotations about multiple axes providing 2D or 3D orientational DoFs. Adjusting orientation is particularly crucial for directional antennas, enabling their main radiation lobes to be precisely aimed towards desired signal sources or receivers, or to optimize polarization alignment with the incoming EM waves \cite{shao20246DMA,shao2024Mag6DMA,shao2024discrete,wu2024Rotatable,shao2024distributed,shao2025hybridnear,ipa,2025polarized,passive,shi20246DMAcellfree,Kinem,aerial6d,		zheng2025rotatable,zhengtian2025rotatable,xiong2025intelligent,zhou2025rotatable,xie2025thz,dai2025rotatable,dai2025rotatableSecure,dai2025demo,wang2025uav}. The antenna orientation/boresight reconfiguration can be achieved by mechanically driven or electronically driven mechanisms~\cite{wu2024Rotatable,zheng2025rotatable,zhengtian2025rotatable}. While mechanically driven methods typically provide a wider control range, electronically driven approaches offer significantly faster response times and better compatibility with existing platforms.

	\paragraph{Implementation Methods}
	Beyond antenna structure design, the choice of implementation methods is also critical and is reviewed as follows.
	
	\textit{Mechanical-based Methods}: Mechanical actuation is a common approach, employing external mechanical structures equipped with actuators such as electric motors or precision gears \cite{zhu2023MAMag}. 6DMAs can be directly mounted on motor-driven shafts \cite{shao2025tutorial,ning2024movable} or guided by linear actuators, providing controlled and often precise changes in their 3D position and orientation. The speed of such motor-based systems typically results in response times ranging from milliseconds to seconds \cite{ning2024movable}. A more compact and potentially faster form of mechanical actuation is achieved through MEMS \cite{marnat2013new}. MEMS technology allows for the fabrication of miniature mechanical components that can physically move or tilt parts of the antenna element. Due to their small scale, MEMS-based MAs can offer faster response times, potentially in the order of microseconds to milliseconds, and thus are suitable for applications requiring rapid, small-scale adjustments, such as implementing flip modes or fine-tuning positions.
	
	\textit{Liquid-based Methods}: Liquid-based methods provide an alternative means of achieving element movement by leveraging the flow characteristics of conductive or dielectric fluids within a confined structure or channel, which can be driven by a syringe \cite{Morishita2013Liquid}, a nano-pump \cite{Shen2021Reconfigurable}, or electrowetting \cite{Wang2022Continuous}. The movement of the fluid can be induced by various means. For example, pressure can be applied using a syringe-like mechanism \cite{Morishita2013Liquid}, or more precise control can be achieved with micro-pumps or nano-pumps that displace the liquid metal within an air chamber or microfluidic channel \cite{Morishita2013Liquid, Shen2021Reconfigurable}. Another technique is electrowetting, where an applied electrical voltage alters the surface tension of the liquid metal in contact with an electrode, generating forces that cause the liquid to move, thereby changing the antenna's effective radiating structure or position \cite{Wang2022Continuous, wu2024fluidMag}. Liquid-based MAs typically enable 1D positional changes, and their response times are generally in the order of milliseconds to seconds \cite{ning2024movable}.
	
	\textit{Pinching Antenna-based Methods}: Pinching antenna-based methods provide a novel way to realize antenna element movement by controlling the radiation point along a waveguide rather than physically moving the entire structure \cite{suzuki2022pinching}. In this architecture, a radio frequency signal is fed into a dielectric waveguide. By bringing a separate dielectric material into close proximity to the waveguide at an arbitrary point, a ``pinch'' is created as an antenna radiator. This effectively creates a movable radiating element whose position can be dynamically controlled by moving the point of the pinch along the waveguide. This technique, first introduced in 2022, enables long-distance antenna positioning along a defined path \cite{suzuki2022pinching}, and it has since been applied to various communication scenarios \cite{ding2024flexible}.
	
	\begin{table*}[t]
		\caption{\label{tab:Key-Characteristics-of-MA}Key Characteristics of Representative
			MA Implementation Methods.}
		
		\centering{}%
		\scalebox{0.9}{
			\begin{tabular}{|c|c|c|c|c|}
				\hline 
				Implementation Methods & Tuning Type & Tuning Range & Tuning Speed & Power Consumption  \tabularnewline
				\hline 
				Motor & Continuous/Discrete & Wide & Moderate (ms$\sim$s) &  High \tabularnewline
				\hline 
				MEMS & Continuous/Discrete & Small & Fast (µs$\sim$ms) & Low \tabularnewline
				\hline 
				Liquid & Continuous & Constrained by linear fluidic channels & Moderate (ms$\sim$s) & Moderate \tabularnewline
				\hline 
				Pinching Antenna & Continuous/Discrete & Constrained by linear waveguides & Moderate (ms$\sim$s) & Moderate \tabularnewline
				\hline 
				Inflatable Structure & Discrete & Fixed (Stowed-to-deployed) & Slow (s) & High \tabularnewline
				\hline 
				Foldable Structure & Discrete & Fixed (Stowed-to-deployed) & Slow (s) & High \tabularnewline
				\hline 
		\end{tabular}}
	\end{table*}
	
	\paragraph{Design Challenges and Possible Solutions}
	The development of movable antenna-elements presents several design challenges \cite{shao2025tutorial,ning2024movable}. For mechanically movable elements, the complexity of the mechanism, the need for ongoing maintenance, and the physical limitations of the actuators (such as size, power, and speed) are primary concerns. Ensuring reliable RF connections and power delivery to the moving element without hindering its range of motion or introducing signal degradation requires careful design of wiring and feed networks. EM coupling between antenna elements can also be affected by their physical movement, necessitating thoughtful placement and potentially adaptive compensation. Furthermore, the energy efficiency of the movement mechanism and the latency incurred during repositioning are critical performance aspects.
	
	Liquid-based elements also face unique challenges. The selection of a suitable fluid material is important, as it must meet stringent criteria related to cost, safety, long-term physical and chemical stability, appropriate melting and evaporation points, and suitable viscosity for controlled flow \cite{ning2024movable}. The EM properties of the fluid, including its permittivity, permeability, conductivity, and loss tangent, are crucial for antenna performance and must be optimized. Achieving complex movements, such as multi-dimensional positioning or rotational functionality for omnidirectional coverage, can be difficult with liquid-based approaches, which are often better suited for simpler linear displacements. Reliable containment and sealing of the liquid material are also important practical considerations to prevent leakage and ensure operational integrity. Solutions to these challenges are pursued through ongoing research into novel materials, advanced actuator designs, sophisticated control algorithms, and durable, flexible interconnect technologies.

	\subsubsection{Movable Antenna-Array}
	Movable antenna-array architectures extend the concept of physical movement to groups of antenna elements. This can involve the coordinated repositioning or reorientation of multiple antenna elements that form an array or subarray, allowing for changes in the array's overall geometry, aperture, or pointing direction. In this subsection, we review the current progress of movable antenna-array, with a focus on their classification, implementation methods, and key design challenges \cite{zhu2025tutorial,shao2025tutorial}.
	
	\paragraph{Classification}
	The classification of array-level movement encompasses several distinct approaches, including arrays composed of individually movable elements and arrays composed of movable subarrays.
	
	\textit{Array of Movable Elements}: An array of individually movable elements represents the most flexible configuration, where each element within the array retains the ability to be moved independently, as described for single movable elements. This offers the maximum possible DoFs for dynamically reconfiguring the array geometry. 
	
	\textit{Array of Movable Subarrays}: A more constrained but often more mechanically feasible approach is the array of movable subarrays \cite{yichi2024movable}. In this configuration, the entire antenna array is typically divided into several smaller, rigid subarrays. These subarrays are then mounted on mechanical systems, such as linear tracks, that allow them to move relative to each other or to a fixed frame. This allows for the adjustment of distances between subarrays, effectively changing the array's baseline and sparsity, which can be used to enhance the effective aperture or modify the beamforming characteristics \cite{Basbug2017design}.
	
	\paragraph{Implementation Methods}
	The implementation of array-level movement can be classified into the following categories.
	
	\textit{Mechanical-based Methods}: Mechanical-based architectures for array-level movement involve altering the geometry of antenna arrays or subarrays using external mechanical systems. One possible approach is the sliding array \cite{yichi2024movable,ning2024movable}, which is composed of one or more subarrays that move along predefined tracks or paths. Another common implementation is the rotatable array, where an entire antenna array is mounted on a platform that permits flexible rotation in terms of yaw, pitch, and roll \cite{shao2024Mag6DMA, shao20246DMA, shao2024discrete}. This architecture is particularly beneficial for installations such as BSs, as it enables the main radiation lobe to be precisely directed toward clusters of users \cite{shao2024Mag6DMA, ning2024movable, shao20246DMA}. 
	
	\textit{Inflatable Structure-based Methods}: Inflatable structures represent a type of deployable array designed to change their physical form from a compact state to an expanded and operational configuration \cite{babuscia2013inflatable}. These arrays utilize internal gas pressure to achieve their intended shape and structural rigidity. Inflatable arrays are especially useful in scenarios where a large aperture is required but stowage volume is a critical constraint. This makes them highly suitable for space applications where they can be compactly stored during launch and expanded once in orbit.
	
	\textit{Foldable Structure-based Methods}: Similar to inflatable structures, foldable arrays are a form of deployable array that can transition between a compact, stowed state and a larger, operational geometry \cite{venkatesh2022origami,Chahat2017deployable}. These arrays often employ origami principles to facilitate their transformation. This adaptability is advantageous for terrestrial systems; for instance, a deployable array on a BS could be folded to minimize wind resistance during severe weather conditions. While offering a cost-effective solution for implementing MAs, the range of motion in foldable arrays is inherently limited by the mechanical properties of the structure. A performance summary table is given
	in Table \ref{tab:Key-Characteristics-of-MA}.
	
		\begin{table*}[t]
		\caption{\label{tab:Comparison-of-RA-MA}Comparison of RA and MA.}
		
		\centering{}
		\scalebox{0.9}{
			\renewcommand{\arraystretch}{1.5}
			\begin{tabular}{|c|c|c|c|c|}
				\hline
				& \multicolumn{2}{c|}{Performance Metrics} & \multicolumn{2}{c|}{Functionality Comparison} \\ 
				\hline
				& Similarities & Differences & Similarities & Differences \\ 
				\hline
				RA & 
				\multirow{2}{*}{\makecell{\\ Power consumption,\\ Integration complexity}} & 
				\makecell{Bandwidth, Frequency,\\ Radiation pattern,\\ Polarization modes} & 
				\multirow{2}{*}{\makecell{RA pattern/polarization\\ reconfiguration vs. 6DMA rotation;\\ Displaced phase centre\\ antennas vs. MA translation;\\ Aim to improve system-level performance}} & 
				\makecell{Frequency\\ reconfiguration} \\ 
				\cline{1-1} \cline{3-3} \cline{5-5}
				MA/6DMA & 
				& 
				\makecell{Movement speed,\\ Movement range,\\ Movement accuracy} & 
				& 
				\makecell{Large-scale channel\\ reconfiguration} \\ 
				\hline
		\end{tabular}}
	\end{table*}

	\paragraph{Design Challenges and Possible Solutions}
	The implementation methods for movable antenna-arrays are predominantly mechanical. For sliding and rotatable arrays, systems involving motors, gears, and linear actuators are commonly used. For example, wheel-and-gear mechanisms or rod-and-motor systems can drive the movement of subarrays along predefined tracks for sliding functionality, or rotate the entire array structure for orientation changes \cite{ning2024movable, shao2024Mag6DMA}. These are essentially scaled-up versions of the actuators used for individual element movement, designed to handle the larger masses and forces involved. Foldable structures rely on integrated mechanical linkages and hinges that allow the array to be collapsed or expanded. The precision of these mechanisms determines the accuracy of the deployed array geometry. Inflatable structures are deployed by filling a flexible envelope with gas, where the antenna elements are typically embedded in or attached to this envelope. The structural integrity and shape are maintained by internal pressure.
	
	Designing and implementing movable antenna-arrays presents considerable challenges. For sliding arrays, the mechanical systems must be robust and reliable over many cycles of movement, and capable of precise control to achieve the desired subarray positions \cite{ning2024movable}. The structural complexity can be significant, and managing the energy consumption and latency associated with moving potentially large subarrays are important design factors. Maintaining consistent and low-loss RF connections to the sliding subarrays as they move is also a critical engineering task, often requiring specialized flexible cables or non-contact power and signal transfer methods. {In addition to specialized flexible cables, non-contact feeding mechanisms have also been considered as a promising solution for maintaining reliable RF connections in MA-arrays. Representative approaches include waveguide-based feeding as well as near-field capacitive or inductive coupling, which can eliminate physical cable connections and simplify the hardware architecture for realizing antenna movement.} Rotatable arrays also face challenges related to the mechanical wear and tear of rotating parts, especially for continuous or frequent adjustments. Ensuring stable and reliable RF connections through rotary joints or slip rings is crucial to prevent signal degradation. Furthermore, even with a fixed array position, the optimization of array orientation in dynamic multiuser communication scenarios can be a complex problem requiring advanced control strategies \cite{zheng2025rotatable,zhengtian2025rotatable}.
	
	Realizing deployable arrays, including foldable and inflatable structures, also presents significant challenges. The achievable range of positions and orientations, as well as the precision of the final deployed geometry, are usually constrained by the mechanical properties of the materials and the design of the deployment mechanism \cite{ning2024movable}. It is essential to ensure the durability and reliability of these mechanisms, particularly for repeated deployment and retraction cycles. Inflatable structures specifically face issues such as maintaining the correct internal pressure, vulnerability to punctures or leaks, and the potential for deformation due to environmental factors like wind or temperature changes. Achieving and maintaining high precision in the shape of a large inflatable antenna surface can be particularly demanding. Addressing these multifaceted challenges requires interdisciplinary solutions involving advanced materials science, precision mechanical engineering, robust control systems, and durable RF interconnects, all tailored to the specific operational requirements and environmental conditions of the movable antenna-array systems.

	\subsection{Comparison of RA and MA}
	\label{subsec:arch_comparison}
	RA and MA/6DMA both offer enhanced DoFs for wireless systems compared to conventional antenna systems, but through fundamentally different mechanisms. RA primarily alters its internal operational characteristics, while MA/6DMA modifies its external spatial properties, specifically its physical position and/or orientation. Understanding their distinct performance metrics and functional capabilities is crucial for leveraging them effectively in future wireless communication and sensing systems.
	
	\subsubsection{Performance Metrics}
	The metrics used to evaluate the performance of RA and MA/6DMA reflect their distinct modes of operation and the parameters they influence. For RA, key performance metrics relate to the ability to adapt the EM properties. \textit{Bandwidth} and tunable \textit{operating frequency range} are critical, indicating the antenna's adaptability to various communication standards or its ability to perform cognitive radio functions by selecting less congested channels. The \textit{radiation pattern characteristics}, such as beamwidth, directivity, side lobe levels, and nulling depth, along with the speed and range of beam steering or shaping, define its radiation reconfigurable capabilities. The range and switching speed between different \textit{polarization modes} (e.g., linear, circular, and their orientations) are important for mitigating polarization mismatch and enhancing diversity. From a system perspective, the \textit{power consumption} associated with the reconfiguration mechanism (e.g., power for PIN diodes, MEMS switches, or tuning circuits) and the \textit{integration complexity}, including the control circuitry and footprint, are significant practical considerations. The efficiency of the antenna in its various states and the losses introduced by reconfiguration components also play a vital role.
	
	For MA/6DMA, the most important performance metrics are the physical displacement capabilities. The \textit{movement speed} is a crucial metric, which dictates how quickly an antenna can change its position or orientation to adapt to dynamic channel conditions or track mobile users. The \textit{movement range}, defining the spatial region within which an antenna can be repositioned, or the angular range through which it can be rotated, determines the spatial DoFs that can be exploited. \textit{Movement accuracy and repeatability} are also vital, ensuring that the antenna can be precisely positioned or oriented to the desired state. Similar to RAs, the \textit{power consumption} of the actuation mechanisms (e.g., motors, MEMS actuators for MA) is a key factor, especially for energy-constrained devices. The \textit{integration complexity} of the mechanical actuators, control systems, and the physical space required for movement, as well as the reliability and durability of these mechanical parts, are also critical for practical deployment. The overhead in terms of time and energy for movement and for acquiring CSI over the entire movement region also impacts system efficiency.
	
	\subsubsection{Functionality Comparison}
	The distinct mechanisms of RA and MA/6DMA lead to different and complementary functional capabilities for enhancing system-level performance in both wireless communication and sensing applications. RA can achieve functionalities like radiation pattern reconfiguration and polarization reconfiguration by altering its internal structure or current distributions. For instance, an RA can steer its main beam towards a user or place nulls towards non-intended receivers electronically, or switch its polarization state to better match the incoming wave. These capabilities are similar to the orientation adjustments of a 6DMA, but are achieved through different mechanisms. While an RA might electronically tilt its beam, a 6DMA physically rotates its entire structure to achieve a similar effect in terms of aligning its directional gain. The speed of reconfiguration is a key factor for RA pattern/polarization reconfiguration or 6DMA rotation. Specifically, RAs typically offer much faster electronic-speed adjustments compared to the mechanical-speed rotations of 6DMA. However, 6DMA might offer a wider continuous range of angular adjustments in 3D space \cite{shao2024Mag6DMA}.
	
	Moreover, RA employing a displaced phase center can achieve effects similar to those of translational movement in MA. By electronically altering its operational mode, an RA can shift the position of its effective radiating center \cite{ning2024movable}. This contrasts with the physical translation of MAs, where the antenna element is physically moved to a new position. Through physical translation, MAs can explore a continuous or finely discretized range of positions within their movement region, potentially accessing a significantly larger set of spatial channel variations than that via phase center displacement in RAs. Phase center displacement in RAs is generally constrained by the antenna's electrical size and design, whereas MAs, though limited by mechanical constraints, can enable larger-scale physical translations \cite{ning2024movable,fu2025extremely}.
	
	In summary, both RA and MA/6DMA technologies aim to enhance the system-level performance of wireless communication and sensing. In communication systems, this enhancement may take the form of increased channel capacity, improved link reliability, enhanced coverage, and more effective interference mitigation. In sensing applications, the objectives typically include improved accuracy, higher resolution, and higher detection probability. RA can achieve these benefits by dynamically adapting its EM response to the environment, whereas MA/6DMA is realized through physical translation and/or orientation to achieve more favorable channel conditions. The choice between RA and MA/6DMA technologies depends on the specific application requirements, the characteristics of the wireless environment, and practical constraints such as speed, power consumption, system complexity, and cost. For example, RA may be preferred for rapid adaptation in highly dynamic interference scenarios, while MA/6DMA may prove more effective in slowly varying environments where larger-scale spatial adjustments can deliver substantial long-term performance gains. The comparison of RA and MA/6DMA is summarized in Table \ref{tab:Comparison-of-RA-MA}.
	
	A promising future direction is the development of hybrid architectures that integrate both RA and MA functionalities. Such systems could leverage the rapid electronic reconfigurability of RA for fine-grained and instantaneous channel adaptation, together with the large-scale spatial optimization of MA for long-term performance enhancement. For instance, an antenna array can be moved to achieve favorable large-scale channel conditions, while individual RA elements within the array dynamically adjust their radiation patterns to track fast fading or suppress time-varying interference. The joint optimization of these complementary DoFs thus represents a fertile area for future research.
	
	\section{RA and MA for Wireless Communications}
	\label{sec:comms}
	
	\subsection{RA for Wireless Communications}\label{Sec:Intro}
	
	By dynamically reconfiguring key EM properties such as radiation pattern, operating frequency, and polarization, RAs offer substantial performance benefits for wireless communication systems. These benefits include improved spectral and energy efficiency, enhanced spectrum utilization, as well as increased link reliability and data rate \cite{Book}.

	\subsubsection{Performance Advantages}
\begin{figure}[!t]
		\centering
		\includegraphics[width = \columnwidth]{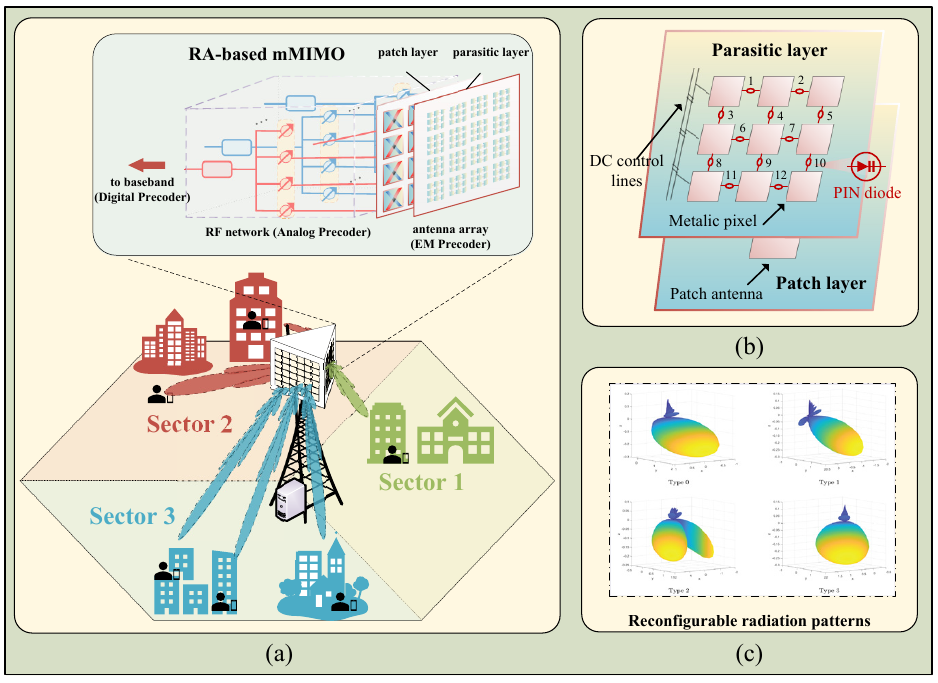}
		\caption{Schematic diagram of pattern-RA-based mMIMO systems: (a)~multiuser downlink transmission and the corresponding mMIMO architecture, (b)~structure of a single pattern-RA, and (c)~examples of 3D radiation pattern produced by a pattern-RA \cite{KeMag}.}
		\label{Array} 
		\vspace*{-5mm}
	\end{figure}

\begin{figure}[!t]
		\centering
		\includegraphics[width = \columnwidth]{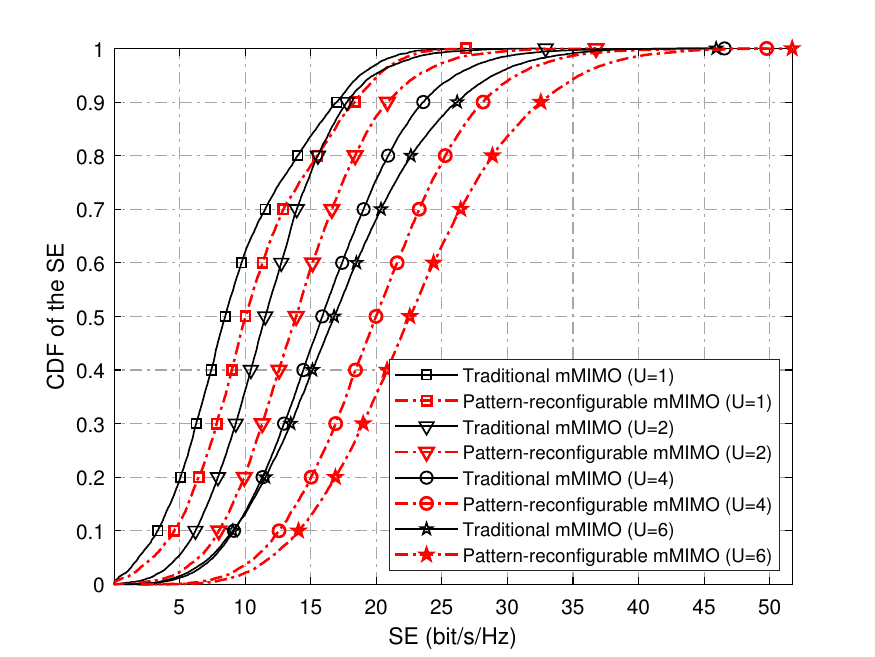}
		\caption{A comparison of CDF curves of traditional mMIMO and pattern-reconfigurable mMIMO under different number of UEs, $U$ \cite{KeMag}.}
		\label{CDF} 
		\vspace*{-5mm}
	\end{figure}

	RA can offer distinct performance advantages across various RA types. Pattern-RAs provide additional DoFs for manipulating antennas at the EM level, significantly enhancing spectral efficiency (SE) and energy efficiency in existing architectures by aligning radiation patterns with propagation channels \cite{HW_Shanghai, HW_France, HKUST_AP, KeMag}. In \cite{KeMag}, a three-level precoding framework was introduced to integrate pattern-RAs with hybrid analog/digital arrays.    
	Fig.~\ref{Array}(a) presents the schematic diagram and hardware design of an RA-based mMIMO system. Unlike traditional mMIMO systems, RA-based hybrid mMIMO systems incorporate an additional parasitic layer atop the patch layer, allowing each antenna to independently shape its radiation pattern. Fig.~\ref{Array}(b) illustrates a single reconfigurable patch antenna (RPA), comprising a patch layer that hosts the patch antenna and a parasitic layer made up of interconnected metallic pixels. Fig.~\ref{Array}(c) illustrates examples of radiation patterns from a single RA, showcasing different radiation directions and shapes. 
	By properly optimizing the PIN connections via offline methods such as genetic algorithms \cite{Genetic}, a desired set of radiation patterns can be selected and maintained for data transmission. Fig.~\ref{CDF} compares the cumulative distribution functions (CDFs) of the SE between traditional mMIMO and pattern-reconfigurable mMIMO architectures, with the system setup detailed in \cite{KeMag}. The results show that for a varying number of UEs, the pattern-reconfigurable mMIMO system consistently achieves higher SE. Within a given predefined set shown in Fig.~\ref{Array}(b), by optimizing radiation pattern combinations at different antennas, the average SE gains of pattern-reconfigurable mMIMO significantly exceed those of its traditional counterpart across different numbers of user equipments (UEs), denoted by $U$, demonstrating the benefits of increased DoFs provided by RAs.
	
	Frequency-RAs are effectively used in cognitive radio systems \cite{wideband_FRA}. These systems are designed to reduce spectrum congestion and improve spectrum utilization by detecting unoccupied or idle frequency bands and dynamically adjusting operational settings to ensure reliable communication. Frequency-RAs can alter their resonant frequency to support multiple wireless services, including WLAN, Bluetooth, and global positioning system (GPS), across a wide range of frequency spectrum. 
	With the increasing availability of spectrum in future 6G communication systems, frequency-RAs offer an efficient solution to carrier aggregation in mmWave and THz communications, enabling simultaneous transmission and reception across multiple frequency channels to enhance data rates. Moreover, frequency-RAs can also be employed in covert communication systems, facilitating channel randomization through frequency hopping \cite{Filtenna, wideband_FRA}.
	
	Polarization-RAs can be used to create diversity for a more reliable link with higher data throughput. Rather than relying on multiple antennas to receive different polarization states, a single antenna with adjustable polarization can be utilized, thereby reducing the size of communication systems for devices with size constraints. Additionally, the polarization domain can enhance data throughput. In \cite{Polar_XiaoPei}, a polarization modulation scheme was proposed to encode additional information in the axial ratio and tilt angle of elliptic polarization, achieving target data rates for 5G wireless systems with reduced bandwidth or fewer antennas.

	\subsubsection{Mode Design and Selection}
	The introduction of additional antenna DoFs necessitates corresponding optimization methods for mode design and selection, aligning with transmission objectives such as enhancing system data rate. 
	In the context of radiation pattern design and selection for RA-aided communication systems, existing pattern optimization techniques can be categorized into two types: discrete space pattern selection and continuous space pattern design. 
	
	For discrete space pattern selection, the goal is to identify optimal combinations of predefined radiation patterns within mMIMO systems to maximize system performance, requiring efficient search strategies within expansive pattern search spaces. An iterative mode search (IMS) method was developed to optimize the radiation pattern combinations at the transmitter, with its complexity scaling linearly relative to the number of transmit antennas \cite{RPA_BF}. Similar methods have been proposed in \cite{KeMag, HW_Shanghai}, underscoring the substantial potential of improving the system throughput by employing reconfigurable patterns in mMIMO systems. 
	For continuous space pattern optimization, the authors in \cite{XJ_Pre_SU, XJ_Pre_MU} formulated the pattern optimization as a continuous space sampling matrix design problem. In \cite{XJ_Pre_SU}, a sequential optimization framework was introduced to refine pattern design within the continuous pattern space for MIMO arrays. Addressing multiuser downlink precoding, joint optimization of symbol-level precoding in the digital domain alongside pattern design in the EM domain was considered in \cite{XJ_Pre_MU}. Additionally, to expand scenarios to multiuser wideband mMIMO pattern design challenges, a spherical harmonic functions-based orthogonal decomposition method was proposed in \cite{KeTcom}, converting continuous pattern function design into the optimization of projection coefficients over spherical harmonic bases, thereby enabling flexible radiation design with desired DoFs. 
	
	\subsubsection{Channel Estimation/Acquisition}
	The authors in \cite{MRA_CE, XJ_CE} addressed the channel estimation challenges in pattern-RA-based mMIMO systems. Unlike traditional mMIMO systems that involve estimating a single channel matrix, pattern-RA-based mMIMO systems typically exhibit multiple channel states due to their various radiation patterns, resulting in significant pilot overhead. In \cite{MRA_CE}, the Gram-Schmidt process was applied to decompose the radiation patterns into orthogonal basis patterns, effectively decoupling antenna radiation patterns from the surrounding channel environment. This decomposition facilitated the development of a joint channel estimation and prediction scheme, thereby reducing the high overhead associated with estimating channels for multiple antenna radiation patterns. Moreover, the authors in \cite{XJ_CE} proposed a deep learning-based channel extrapolation method. During channel estimation, different patterns were assigned to different antennas at the transmitter, and a deep neural network was employed at the receiver to extrapolate channels for other patterns. This method can achieve a good channel estimation performance with low pilot overhead. However, further investigation into wideband channel estimation with larger sets of radiation patterns is still underexplored \cite{zheng2025tri1,zheng2025tri2}.
	
	\begin{figure*}[!t]
		\centering
		\includegraphics[width=160mm]{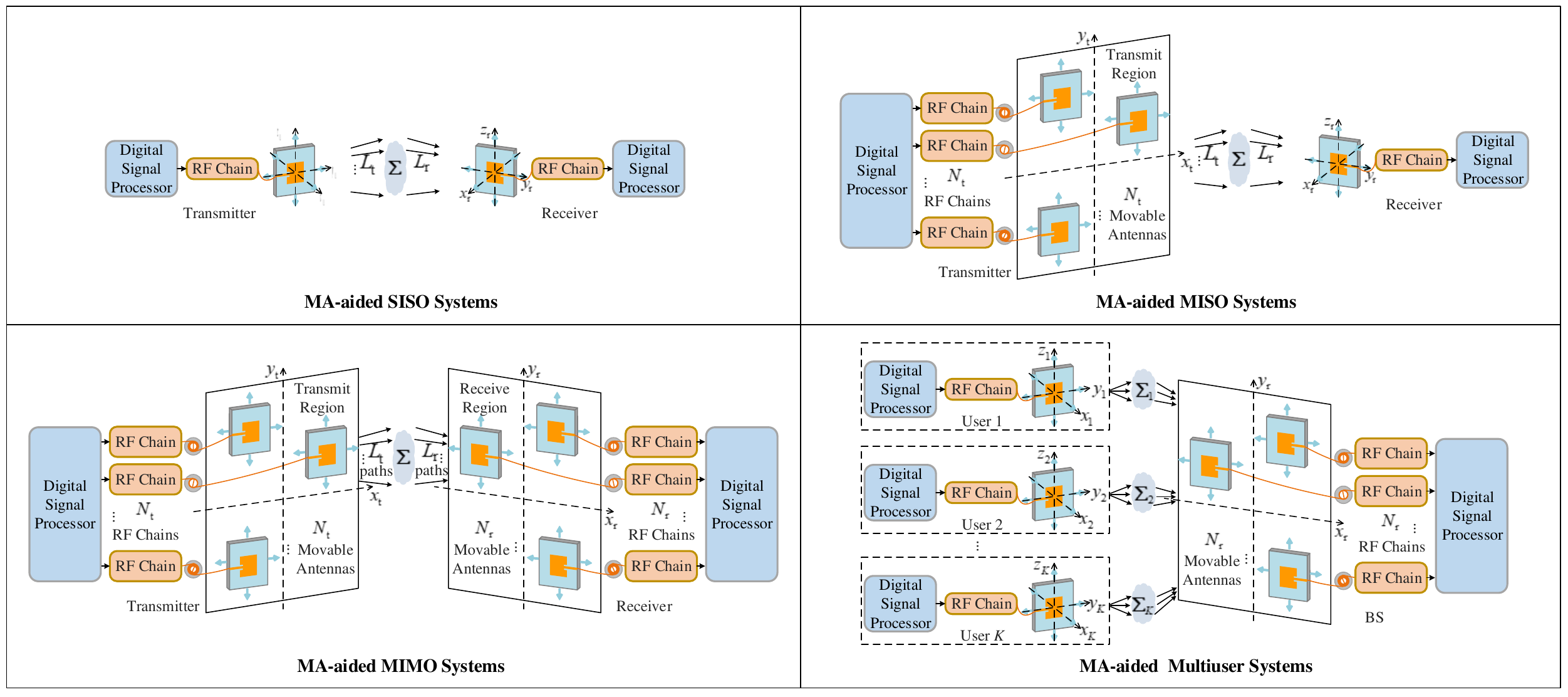}
		\caption{Illustration of MA-aided SISO, MISO, MIMO, and multiuser communication systems. In these systems, the antenna(s) at the transmitter and/or receiver can be moved to improve communication performance.}
		\label{Fig_communication}
	\end{figure*}

	\subsubsection{Extension to IRS-aided Wireless Communications}
	
	RA technology has demonstrated significant capabilities in enhancing communication performance by dynamically adjusting internal antenna parameters such as frequency, radiation pattern, and polarization. However, the active RA's capabilities remain constrained by its localized hardware-centric adaptability, due to the antenna's intrinsic reconfiguration range and physical placement. Challenges such as limited coverage in dense multipath environments and high energy consumption for rapid pattern switching still persist, particularly in non-line-of-sight (NLoS) scenarios.
	
	IRS builds upon the principles of RA by further extending from device-level reconfiguration to environment-level programmability. While the active RA modifies the inherent properties of the transceiver, IRS reshapes the wireless propagation environment, providing a scalable and energy-efficient complement to RA-based transceivers~\cite{zheng2022survey, wu2024ISfor6G,9140329}. Specifically, IRS comprises digitally controlled passive elements capable of precisely manipulating EM wave reflections by adjustable phase shifts of the incident signals~\cite{wu2019IRS,9849035,9039554}. Therefore, IRS can directly reshape the wireless propagation channel to optimize signal transmission, such as enhancing the received signal power for intended users and mitigating interference for unintended users \cite{shi2023intelligent,shi2024secrecy,shi2025combating,IRS25PL}.
	
	To fully exploit the potential of IRS, accurate CSI is essential. In practical applications, the lack of signal processing capabilities in passive IRSs prevents them from directly sensing the wireless environment, thereby making channel estimation a critical and challenging task. One feasible approach is to integrate active sensors into the IRS, enabling it to autonomously acquire CSI~\cite{9335617,9127834,9340586}. For example, a low-complexity method using an L-shaped sensing array on the IRS was proposed to separately estimate BS-IRS and user-IRS channels based on angle-of-arrival (AoA) and path gain~\cite{9335617}.  However, this integration inevitably increases the energy consumption and hardware cost of the IRS. Another method relies on transmitting pilot signals~\cite{9039554,Mishra2019Channel,zheng2020IRSOFDM,9261597,9053695,9195133}, allowing the receiver to directly estimate the transmitter-IRS-receiver cascaded channel. One representative method employs ON/OFF training reflection patterns for distinct IRS elements in different time slots to separate each element's contribution at the receiver~\cite{9039554,Mishra2019Channel}. Other studies focus on increasing estimation accuracy and reducing training overhead by adopting full-ON training reflection patterns, carefully designed pilot sequences, and element-grouping strategy~\cite{zheng2020IRSOFDM,9261597,9053695,9195133}. In addition, low-complexity techniques such as compressed sensing~\cite{9354904,9403420} and codebook-based feedback~\cite{9367208,9366894} can also be applied to reduce the training overhead for channel estimation. On the other hand, to achieve the theoretical performance gain, precise control of IRS beamforming/reflection is required to achieve high channel gain~\cite{8970580}. By leveraging the aforementioned channel estimation methods, joint optimization of IRS passive reflection and BS active beamforming can be performed using the estimated CSI. However, acquiring perfect CSI remains challenging due to channel aging, constrained training and feedback capacity, and the presence of noise. This challenge is further exacerbated in IRS-aided systems given the need to estimate additional IRS-associated channels. Consequently, extensive research has focused on IRS passive beamforming/reflection designs under imperfect or statistical/hybrid CSI~\cite{9110587,9352530,9424713,9423667}. Alternatively, some works seek to bypass explicit CSI acquisition by employing beam training, deep learning, and other data-driven approaches~\cite{9129778,9513283,9206080,sun2024power,10907801}. These efforts collectively aim to balance performance and complexity in IRS-aided systems under realistic CSI constraints.
	
	Besides, the performance of IRS systems is also highly dependent on deployment strategies. In static or quasi-static wireless environments, IRS can be strategically deployed at fixed locations to enhance coverage and bypass blockages. However, in highly dynamic environments, static IRSs may be insufficient to adapt to rapid environmental changes. To address this limitation, mobile IRSs mounted on vehicles or UAVs have been proposed, enabling flexible deployment and real-time adaptation to user mobility and network dynamics~\cite{9661068,9973357,10028753}. These mobile configurations introduce new design dimensions, such as joint trajectory and reflection optimization, enabling IRS to maintain consistent performance even in scenarios with high mobility or uncertain propagation conditions. In addition, prior works have mainly focused on single-IRS deployments located near the user, BS, or relay to enhance coverage or achieve comparable performance to mMIMO~\cite{9745477,10536035,9464248,9866003}. However, such setups face challenges including limited reflection coverage, blockage susceptibility, constrained beamforming gains, and low spatial multiplexing due to channel correlation. To overcome these limitations, recent studies have focused on double-/multi-IRS deployments, enabling coordinated reflection to enhance link robustness and system performance~\cite{9849035,9362274,9241706,9373363,10536035,9448236,FU2025MultiIRS,FU2024MultiIRS}. For instance, the authors in~\cite{9362274} extended the alternating optimization based joint active/passive beamforming design to double-IRS systems by employing semidefinite relaxation and bisection methods to efficiently solve the max-min signal-to-interference-plus-noise ratio (SINR)/rate optimization problem. Furthermore, by exploiting multi-reflection paths across distributed IRSs, multi-IRS systems can bypass obstacles, enhance diversity, and improve beamforming efficiency, thereby supporting more flexible user association and quality-of-service provisioning~\cite{9448236}.

	As a further advance, the co-deployment and integration of IRS with RA technologies has the potential to synergistically enhance system flexibility and efficiency. Such integration combines RA's dynamic internal parameter adjustment with IRS's ability to manipulate the external propagation environment, creating more robust and adaptable wireless networks. Nevertheless, substantial technical challenges remain, particularly in accurate cascaded channel estimation, practical beamforming design, and managing the complexities associated with realistic hardware constraints. Therefore, continued research is essential to address these challenges, enabling the effective integration of IRS and RA technologies to fully unlock their combined potential in future wireless communication systems.

	\subsection{MA for Wireless Communications}
	\label{subsec:comms_ma}
	MA/6DMA represents a significant paradigm shift in wireless communication system design, offering the potential to substantially enhance performance by dynamically altering antenna positions and/or orientations. In contrast to traditional FPA systems, which are unable to exploit spatial DoFs through antenna position optimization, MA/6DMA systems can actively reconfigure their array geometries to leverage more favorable channel conditions. This section explores the performance benefits, challenges in channel modeling, techniques for channel acquisition, and movement strategies associated with MA/6DMA-aided wireless communication systems.
	
	\subsubsection{Performance Advantages}
	The fundamental performance gain of MA stems from its ability to exploit the spatial variations inherent in wireless channels \cite{zhu2022MAmodel}. By physically moving the antenna elements, an MA system can adjust the phases of multipath components to achieve constructive superposition at the receiver, thereby maximizing received signal power, or destructive superposition to null interference. This capability introduces new DoFs for obtaining favorable channel conditions. As shown in Fig.~\ref{Fig_communication}, in MA-aided SISO systems, the primary advantage lies in improving the effective channel gain. By optimizing the antenna position, the phases of multiple channel paths can be aligned, leading to a significant increase in received signal power compared to an FPA that might be in deep fading \cite{zhu2022MAmodel}. Conversely, MAs can also be positioned to minimize the channel gain from an interfering source. Moreover, for wideband systems, while frequency selectivity makes it harder to align all paths across all subcarriers simultaneously, substantial average channel power gains are still achievable via antenna position optimization \cite{zhu2024wideband}.
	
	For MA-aided MISO or single-input multiple-output (SIMO) systems, antenna movement offers benefits beyond simple signal power enhancement. When MAs form an array, their positions can be jointly optimized to reshape the array geometry \cite{zhu2023MAarray, ma2024multi, wang2024flexible}. Under pure line-of-sight (LoS) conditions, this allows for highly flexible beamforming, such as achieving full array gain while steering nulls towards multiple interference directions (i.e., beam nulling) \cite{zhu2023MAarray}, creating multiple beams towards different users (i.e., multi-beam forming) \cite{ma2024multi}, or ensuring uniform coverage over a wide angular region (i.e., wide-beam coverage) \cite{wang2024flexible}. These capabilities often exceed those achievable by FPAs relying only on antenna weight optimization. In point-to-point NLoS scenarios, MA arrays can optimize antenna positions to maximize the overall received signal power \cite{Hu2024comp, LaiXZ_FAS_RIS}, where each antenna is positioned to enhance its individual signal power, subject to inter-antenna spacing constraints to avoid antenna coupling.
	
	In MA-aided MIMO systems, antenna position optimization can significantly enhance spatial multiplexing performance \cite{ma2022MAmimo}. In low SNR regimes, MAs can be configured to maximize the largest singular value of the MIMO channel matrix, improving single-stream beamforming. In high SNR regimes, positions can be optimized to balance the singular values, catering to water-filling power allocation strategies across multiple eigenchannels \cite{ma2022MAmimo}. This adaptation can be based on instantaneous CSI for slowly varying channels \cite{ma2022MAmimo} or statistical CSI for fast-fading channels to improve ergodic capacity while reducing movement overhead \cite{chen2023joint, Ye2024fluidstc, zheng2024twotimeMA}.
	Moreover, for MA-aided multiuser systems, MAs can enhance performance by not only improving individual user channel gains but also by reducing inter-user interference \cite{zhu2023MAmultiuser, xiao2023multiuser, wu2023movable, Yang2024movable,yang2025flexible,pi2025movable,ding2025energy}. Optimizing antenna positions at the BS or user devices can reduce channel correlations between users \cite{zhu2023MAmultiuser}, thereby enabling more effective spatial separation and leading to improved multiuser communication performance.
	
	\subsubsection{Channel Modeling}
	Accurate channel models are crucial for the design and evaluation of MA systems, as they must capture the continuous variation of the wireless channel with respect to antenna position and orientation. These channel models can be broadly categorized into those based on physical path propagation and those based on statistical properties.
	
	The field-response channel model provides a foundational framework by characterizing the wireless channel based on physical propagation paths \cite{zhu2022MAmodel, ma2022MAmimo, zhu2024nearfield, ding2024near, shao20246DMA, zhu2024wideband, chen2024joint}. This model expresses the channel as a function of the transmitter and receiver antenna positions and/or orientations by considering the superposition of multiple signal paths, each defined by its angle-of-departure (AoD), AoA, and a complex path response coefficient. While initially applied for deterministic modeling, this physically grounded model has recently been extended for statistical channel modeling to characterize random channel variations caused by factors like the local movement of users \cite{yan2025movable, chen2023joint}. The field-response framework is general and applicable to both narrowband and wideband systems, as well as to far-field and near-field propagation conditions. For wideband scenarios, the model is extended by incorporating distinct path responses and field response vectors (FRVs) for each delay tap \cite{zhu2024wideband}. Under near-field conditions, spherical waves are used in place of the planar wave assumption \cite{zhu2024nearfield, ding2024near, chen2024joint}. In 6DMA systems, the path response matrix further depends on antenna orientation matrices, capturing radiation pattern and polarization effects \cite{shao20246DMA, shao2024discrete, shao2024Mag6DMA, pi20246DMAcoordi, zhang20246DMAhybrid}. The main challenge of this model is its analytical and computational complexity, which grows with the number of path parameters involved.
	
	Other statistical channel models, such as those based on spatial correlation, offer an alternative approach \cite{Ozdogan2019MIMOfading, Wang2024cellfree}. These models are valuable for their analytical tractability and can offer robustness to modeling inaccuracies when their statistical assumptions are valid \cite{Khammassi2023approx, Psomas2023continuous, New2024fluid, Espinosa2024block, Rostami2023copula}. Classical models like Jake’s model and its various refinements have been applied to MA-aided wireless systems, especially in rich scattering environments, by employing statistical distributions (e.g., Rayleigh or Rician fading) and spatial-temporal correlation functions to describe the channel's evolution as the antenna moves \cite{Wong2021fluid,zhao2009single}. However, a key limitation is that they often overlook the detailed influence of antenna radiation patterns and deterministic multipath structures. {As such, these statistical models can be viewed as approximations of the more general field-response channel model, particularly under conditions involving a large number of randomly distributed multipath components \cite{zhu2022MAmodel}.}

	\subsubsection{Channel Acquisition}
	To fully exploit the spatial DoFs via antenna movement optimization, it is essential to acquire instantaneous/statistical CSI between any two points within the transmitter and receiver antenna moving regions in MA-aided wireless communication systems \cite{ma2023MAestimation, xiao2023channel, Xu2024estimation}. This process involves constructing a channel mapping over the antenna moving regions. {The choice of channel acquisition methods is closely related to the employed channel model. Model-based channel acquisition methods are particularly suited for the field-response channel model because they exploit the underlying physical structure to estimate path parameters. In contrast, model-free channel acquisition methods are typically applied when a deterministic channel structure is unavailable or when statistical models are used.}
	
	Model-based channel acquisition techniques rely on the field-response channel model. The main objective is to estimate the underlying field response information (FRI), which consists of wave vectors determined by AoAs/AoDs as well as the path response matrix (PRM), using a limited number of channel measurements at selected antenna positions. Leveraging the inherent angular sparsity of dominant propagation paths \cite{ma2023MAestimation, xiao2023channel, zhang2024TensorCE,xiao2024channelwide}, compressed sensing techniques (such as orthogonal matching pursuit (OMP) \cite{lee2016channel}) can be employed to recover the FRI based on the channel measurements at selected antenna positions \cite{wei2025super}. An alternative model-based approach leverages tensor decomposition \cite{zhang2024TensorCE}. By arranging the MA measurement positions in a uniform planar array (UPA) configuration, the received pilot signals can be structured into a multi-dimensional tensor. This tensor can then be processed using decompositions such as the canonical polyadic decomposition to extract wave vector information. This method enables grid-free and super-resolution estimation of wave vectors, particularly effective in high-SNR scenarios. These model-based methods significantly reduce measurement overhead, as the number of parameters to estimate depends on the number of dominant paths rather than the size of the movement region. However, the accuracy of these methods is highly dependent on the validity of the underlying channel model and can degrade significantly in the presence of model mismatches or measurement noise.
	
	On the other hand, model-free channel acquisition methods do not rely on a predefined channel structure. Instead, they involve directly sampling the channel at discrete positions and interpolating or predicting the channel at unmeasured locations \cite{Skouroumounis2023fluidCE, new2024CE, wang2023FACE, ji2024correlation, zhang2024MLCE,jang2025new,huang2025cnn, cui2024near,zhang2025deep}. Basic interpolation techniques may set the channel at an unmeasured position to be equal to that of its nearest measured position \cite{Skouroumounis2023fluidCE, new2024CE}. Moreover, the Bayesian linear regression method treats the channel as a Gaussian random field and uses channel measurements at discrete positions to update a posterior distribution over the entire channel map \cite{zhang2023successive, cui2024near}. Additionally, machine learning-based methods have been proposed, in which neural networks are trained using datasets of channel measurements at discrete positions to predict channel at unmeasured positions \cite{ji2024correlation, zhang2024MLCE,jang2025new,huang2025cnn}. Model-free methods are more robust to modeling inaccuracies and are particularly effective in low-dimensional or small movement regions. However, since accurate interpolation or prediction of the channel at unmeasured positions typically requires dense (sub-wavelength order) spatial sampling, the measurement overhead of model-free channel acquisition methods generally scales with the size and dimensionality of the movement region.
	
	\subsubsection{Antenna Movement Optimization}
	Practical and efficient antenna movement strategies are crucial for optimizing MA systems, which should be determined based on the type of acquired CSI and the associated time and energy overheads of antenna movement. Channel-based optimization strategies leverage acquired CSI to optimize antenna movement. These strategies aim to optimize antenna positions for objectives such as maximizing received SNR \cite{zhu2022MAmodel, zeng2024csi}, mitigating interference, enhancing MIMO capacity \cite{ma2022MAmimo, chen2023joint}, or achieving flexible beamforming \cite{zhu2023MAarray, ma2024multi, wang2024flexible}. However, the underlying optimization problems are typically non-convex due to the complex and nonlinear relationship between antenna positions and communication objectives. To address this, a variety of algorithms have been explored, including local optimization methods such as gradient-based \cite{zhu2023MAmultiuser, shao20246DMA, peng2024jointISAC, zhu2024wideband, hu2024power} and successive convex approximation (SCA) techniques \cite{ma2022MAmimo, WuHS_MA_RIS_ISAC, ma2024multi, wang2024flexible}, as well as global optimization approaches like particle swarm optimization \cite{xiao2023multiuser, kuang2024movableISAC, DingJZ_MA_FD_secure_1} and graph-theoretic methods \cite{mei2024movable, mei2024movable_secure} tailored for discretized antenna movement regions. Another approach formulates the optimization of antenna positions as a sparse recovery problem, involving creating a dictionary of candidate antenna positions and employing greedy pursuit algorithms to select the subset of positions that maximizes the system objective like the sum-rate \cite{yang2025flexible,pi2025movable}. More recently, AI techniques \cite{zhu2023MAMag, WangC_FAS_RIS_AI_survey} have emerged as promising tools for learning effective antenna movement strategies, even in scenarios with limited or implicit CSI, thereby bridging the gap with model-based training paradigms.
	
	Antenna position optimization can be based on either instantaneous \cite{ma2022MAmimo} or statistical CSI \cite{chen2023joint, shao20246DMA}. Instantaneous CSI-based optimization enables real-time adaptation to prevailing channel conditions and is particularly effective in quasi-static environments with long channel coherence times, where the overhead associated with channel estimation, computation, and mechanical repositioning is manageable. In contrast, in fast-fading environments, the frequent need for CSI acquisition and antenna movement may lead to excessive overhead. In such scenarios, statistical CSI-based approaches are more appropriate \cite{Hu20242024twotimeMA, zheng2024twotimeMA}. These methods optimize antenna positions based on long-term performance metrics (e.g., ergodic capacity) by leveraging stable statistical characteristics of the channel, including AoD/AoA distributions and spatial correlation, thereby reducing the frequency of antenna movement while still exploiting spatial DoFs.
	
	Practical constraints are critical in antenna movement design. Key factors include movement speed, which determines how quick the MA system can be reconfigured; movement time and associated power consumption, which are particularly significant for energy-constrained systems; and the available 1D/2D/3D translation and/or rotation range \cite{Vahid2024movable, Vahid2024MAplan, li2024minimizing, wang2024MAdelay}. Additionally, movement accuracy is essential, as positioning errors on the order of the wavelength can significantly degrade the expected performance gains \cite{yao2024rethinking}. As a result, effective MA system design must achieve a balance between communication performance improvements and the mechanical and energy costs of antenna movement. Achieving this balance often requires customized hardware architectures and intelligent scheduling mechanisms to ensure efficient and reliable operation.

	\subsection{Performance Comparison}
	\label{subsec:comms_comparison}
	While both MA/6DMA and RA aim to enhance wireless system performance by introducing additional DoFs, their fundamentally different operational mechanisms lead to distinct advantages and trade-offs. This section compares their performance under various scenarios, highlighting how their unique characteristics influence system-level performance.
	
	\begin{figure}[!t]
		\centering
		\includegraphics[width=80mm]{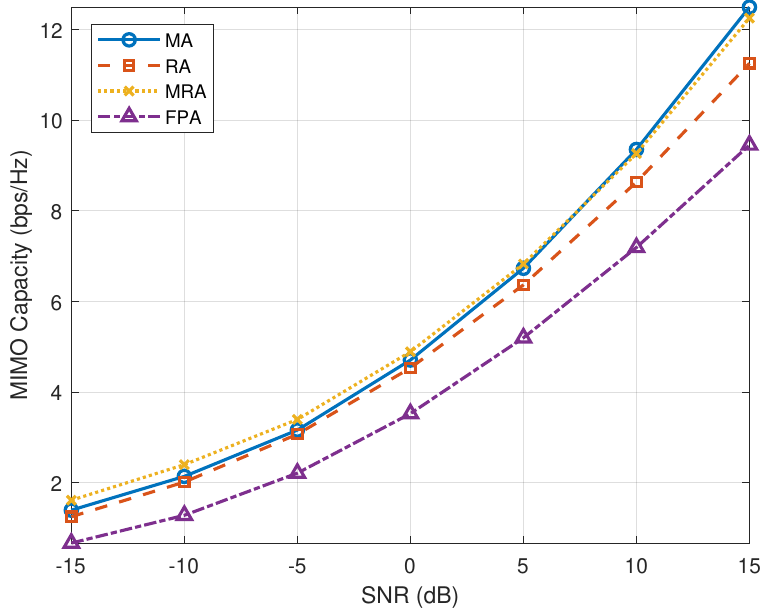}
		\caption{Comparison of MIMO capacity for MA and RA systems versus SNR.}
		\label{Fig_MIMO_rate_SNR}
	\end{figure}
	
	To evaluate and compare the performance of MA-aided and RA-aided wireless communication systems, we show in Fig.~\ref{Fig_MIMO_rate_SNR} the capacity of MA-MIMO, RA-MIMO, movable and reconfigurable antenna (MRA)-MIMO, and FPA-MIMO systems versus the average SNR. All the MIMO systems employ $4$ transmit antennas and $4$ receive antennas. We consider the geometric channel model with $5$ channel paths. The total channel power is normalized as $1$, and the ratio of the average power between LoS and NLoS paths is set as $10$ dB. The AoDs and AoAs for all paths are modeled as independent and identically distributed (i.i.d.) random variables, uniformly distributed over $[0, 150^\circ]$. For the MA and MRA systems, the antennas can be moved within a square region of size $3\lambda\times3\lambda$, where $\lambda$ is the wavelength. A minimum distance of $\lambda/2$ is required between any two antennas at the same side to avoid coupling. The four systems are configured as follows: (i) MA: The antennas have a fixed isotropic radiation pattern. Their positions are iteratively searched within the given movement region to maximize the MIMO capacity; (ii) RA: The transmit and receive antennas form two uniform linear arrays (ULAs) with $\lambda/2$ inter-antenna spacing. Each antenna has three candidate reconfigurable directional radiation patterns \cite{wang2025EMally}. The radiation pattern of each antenna is iteratively searched to maximize capacity; (iii) MRA: The antennas have the same reconfigurable pattern as the RA system. A two-stage alternating optimization is performed: First, the antenna positions are iteratively searched with a fixed default radiation pattern; second, using these optimized positions, the antenna radiation patterns are then iteratively searched; (iv) FPA: The transmit and receive antennas form two ULAs with $\lambda/2$ inter-antenna spacing, where each antenna has a fixed isotropic radiation pattern.
	
	As shown in Fig.~\ref{Fig_MIMO_rate_SNR}, the MA, RA, and MRA systems all substantially outperform the conventional FPA system, demonstrating the significant potential of adapting antenna positions or radiation patterns to enhance wireless communication performance. Specifically, the superior performance of MRA over RA reveals that incorporating the additional DoFs of antenna movement into RA systems yields additional capacity gains. Furthermore, the system performance across different SNR regimes reveals fundamental trade-offs between RA and MA systems. In the low-SNR regime, the MRA system outperforms the MA system. This is because MIMO systems typically transmit a single data stream under low-SNR conditions, and both MRA and RA can reconfigure their radiation patterns to enhance the beamforming gain along the LoS path. In contrast, in the high-SNR regime, the MA system achieves superior performance over both RA and MRA systems. This is because the MA system optimizes the positions of its isotropic antennas, allowing it to fully leverage the rich multipath environment from all angular directions to effectively decorrelate channel paths and maximize spatial multiplexing gain. The MRA system, however, is constrained to a limited set of pre-defined directional radiation patterns. This inherent directionality can be a disadvantage for spatial multiplexing, as each antenna focuses on a specific angular sector and becomes less sensitive to signal paths arriving from other directions. Consequently, in this high-SNR scenario where maximizing multiplexing gain is critical, the MA system holds an advantage over the MRA system due to its ability to fully exploit the spatial characteristics of the channel.
	
	\begin{figure}[!t]
		\centering
		\includegraphics[width=80mm]{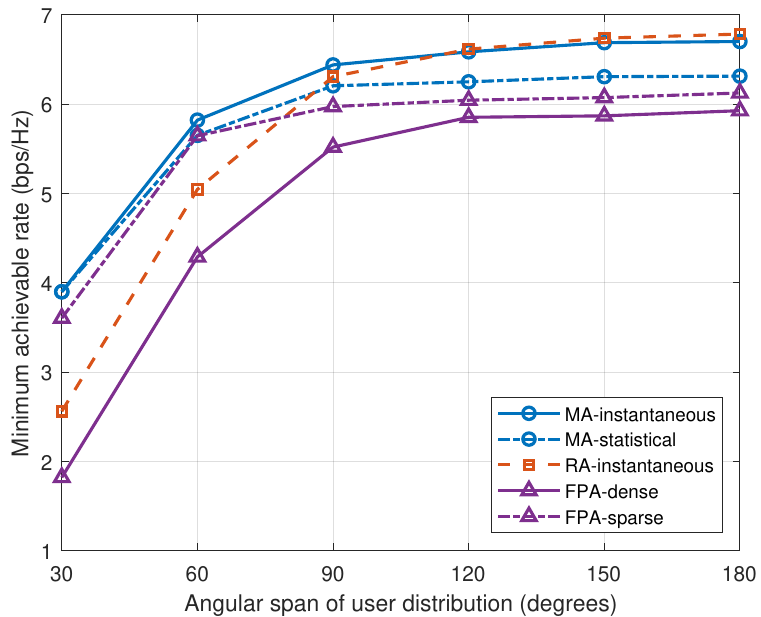}
		\caption{Comparison of minimum achievable rate for MA and RA systems versus angular span of user distribution.}
		\label{Fig_MU_rate_size}
	\end{figure}
	
	To evaluate and compare the performance of MA-aided and RA-aided multiuser wireless communication systems, Fig.~\ref{Fig_MU_rate_size} shows the minimum achievable rate versus the angular span of user distribution. We consider a downlink multiuser system where a BS, equipped with $4$ antennas, serves $2$ single-FPA users. The users are randomly distributed along a circular arc with a radius of $50$ m centered at the BS. The transmit SNR is set to $100$ dB, and the carrier wavelength is $\lambda = 0.05$ m. Zero-forcing (ZF) beamforming is employed at the BS. In the MA schemes, the BS antennas are movable along a linear segment of length $4\lambda$. The five schemes are configured as follows: (i) MA-instantaneous: Antenna positions are iteratively searched for each random user location realization to maximize the instantaneous minimum achievable rate; (ii) MA-statistical: Antenna positions are optimized once for a given angular span by maximizing the average minimum achievable rate over a large number of randomly generated user distributions within that span; (iii) RA: The BS antennas form a ULA with $\lambda/2$ spacing. Each antenna has three selectable directional radiation patterns \cite{wang2025EMally}, which are iteratively selected to maximize the instantaneous minimum achievable rate; (iv) FPA-dense: A fixed ULA with $\lambda/2$ inter-antenna spacing and isotropic radiation patterns; (v) FPA-sparse: A fixed ULA with $2\lambda$ inter-antenna spacing and isotropic radiation patterns.
	
	As shown in Fig.~\ref{Fig_MU_rate_size}, both MA and RA schemes significantly outperform the FPA schemes, which highlights the performance enhancement of adapting the antenna array to user distribution. Additionally, the performance of all schemes generally improves with increasing angular span, as larger user angular separation reduces channel correlation, thereby enhancing the effectiveness of ZF-based spatial multiplexing. Notably, the performance of the MA-statistical scheme closely approaches that of the MA-instantaneous scheme, demonstrating the practical effectiveness of statistical CSI-based antenna position optimization without requiring frequent repositioning. {Furthermore, the MA schemes outperform the RA scheme at smaller angular spans, where users’ channels tend to be highly correlated. In this regime, optimizing MAs' positions can effectively reduce inter-user channel correlation, which is more beneficial than reconfiguring radiation patterns. In contrast, when the angular span becomes larger, users’ channels are inherently less correlated due to increased spatial separation. In this case, the RA scheme can more effectively enhance the received signal strength by reconfiguring distinct directional radiation patterns toward different users, leading to improved achievable rate performance.}
	
	\textit{Lessons Learned}: This section highlights that RA and MA provide complementary advantages. RA enables rapid control of EM properties to achieve energy focusing, whereas MA leverages large-scale spatial channel variations through physical repositioning to enhance spatial decorrelation. This gives rise to a fundamental performance trade-off: RA is well-suited for maximizing beamforming gain in low-SNR or LoS-dominant scenarios, while MA is more effective in maximizing spatial multiplexing and mitigating interference in high-SNR, multipath-rich environments. The primary design challenges for both technologies lie in devising efficient channel acquisition methods and low-complexity optimization strategies to fully exploit their practical potential. Moreover, the integration of RA and MA technologies offers a promising path forward, enabling the joint optimization of their complementary strengths to realize highly adaptive and efficient communication systems.
	
	\section{RA and MA for Wireless Sensing and ISAC}
	\label{sec:sensing_isac}
	
	\subsection{RA for Wireless Sensing and ISAC}
	RA introduces additional DoFs that enhance the performance of wireless sensing and ISAC systems. By dynamically adjusting its EM properties like radiation patterns, a single RA can fulfill functions that conventionally require multiple antennas such as AoA estimation, thereby improving sensing accuracy and enabling more flexible ISAC operations.
	
	\label{subsec:sensing_ra}
	\subsubsection{Reconfigurable Elements-aided Wireless Sensing}
	\label{ssubsec:sensing_ra_elements}
	\begin{table*}[t]\normalsize
		\caption{Summary of RA for Wireless Sensing.}
		\centering
		\label{ra-sensing}
		\vspace{-5mm}
		\begin{threeparttable}
			\begin{center}
				\renewcommand{\arraystretch}{1.1}
				\scalebox{0.9}
				{
					\begin{tabular}{|c|c|c|c|c|}
						\hline
						\diagbox{Sensing Task}{Type} & Pattern-RA & Leaky-wave Antenna  & Sectorized Antenna    \\ \hline
						AoA Estimation &\cite{Tai2005TAP-RA-AOA,Qian2015SPAWC-RA-AOA,Qian2016ICASSP-RA-AOA,Yazdani2017MILCOM-RA-AOA,Kulas2017AWPL-RA-AOA,Kulas2018AWPL-RA-AOA,Tarkowski2019AWPL-RA-AOA,Bshara2021Access-RA-AOA} & \cite{Abiel2011TAP-Leaky-AOA,Paaso2013PIMRC-Leaky-AOA,Paaso2017TAP-Leaky-AOA,Poveda2019TAP-Leaky-AOA,Ning2020CL-Leaky-AOA,Poveda2020Access-Leaky-AOD,Zaka2023IOTJ-Leaky-AOA-CFO}  & \cite{Werner2015JSAC-Sector-AOA,Pell2022PIMRC-Sector-AOA}   \\ \hline 
						Localization & \cite{Rzy2016AWPL-RA-Loc,Groth2023EuCAP-RA-Loc} & \cite{Hakka2014CROWNCOM-Leaky-Loc}   & \cite{Werner2013WONS-Sector-Loc,Wang2014WCL-Sector-Loc,Werner2016TVT-Sector-Loc}  \\ \hline              
				\end{tabular}}
			\end{center}
		\end{threeparttable} 
	\end{table*}
	
	The use of RAs in wireless sensing has been extensively studied, with numerous implementations enhancing capabilities such as AoA estimation and localization \cite{Werner2013WONS-Sector-Loc,Wang2014WCL-Sector-Loc,Werner2015JSAC-Sector-AOA,Werner2016TVT-Sector-Loc,Pell2022PIMRC-Sector-AOA,Abiel2011TAP-Leaky-AOA,Paaso2013PIMRC-Leaky-AOA,Hakka2014CROWNCOM-Leaky-Loc,Paaso2017TAP-Leaky-AOA,Poveda2019TAP-Leaky-AOA,Ning2020CL-Leaky-AOA,Poveda2020Access-Leaky-AOD,Zaka2023IOTJ-Leaky-AOA-CFO,Tai2005TAP-RA-AOA,Qian2015SPAWC-RA-AOA,Rzy2016AWPL-RA-Loc,Qian2016ICASSP-RA-AOA,Yazdani2017MILCOM-RA-AOA,Kulas2017AWPL-RA-AOA,Kulas2018AWPL-RA-AOA,Tarkowski2019AWPL-RA-AOA,Bshara2021Access-RA-AOA,Groth2023EuCAP-RA-Loc,chen2025dbraa,fadakar2025hybrid,chen2025integrated,zheng2025electromagnetically}, as summarized in Table \ref{ra-sensing}. The primary advantage of RA in wireless sensing lies in its ability to equip a single antenna with angular resolution. Two prominent RA-based sensing architectures are pattern-RAs and leaky-wave antennas (LWAs).  
	
	Pattern-RA, often implemented using electronically ESPAR designs, consists of a single active radiating element surrounded by passive parasitic elements. By dynamically tuning the reactance of these passive elements, pattern-RA can steer the radiation beams directionally, enabling \emph{frequency-independent} AoA estimation \cite{Tai2005TAP-RA-AOA,Qian2015SPAWC-RA-AOA,Rzy2016AWPL-RA-Loc,Qian2016ICASSP-RA-AOA,Yazdani2017MILCOM-RA-AOA,Kulas2017AWPL-RA-AOA,Kulas2018AWPL-RA-AOA,Tarkowski2019AWPL-RA-AOA,Bshara2021Access-RA-AOA,Groth2023EuCAP-RA-Loc}. The study in \cite{Tai2005TAP-RA-AOA} represents one of the earliest attempts to apply pattern-RAs for AoA estimation using the multiple signal classification (MUSIC) algorithm. In \cite{Qian2015SPAWC-RA-AOA,Qian2016ICASSP-RA-AOA,Yazdani2017MILCOM-RA-AOA}, compressed sensing frameworks were developed for pattern-RA-enabled AoA estimation by formulating the problem with sparse representation. The work in \cite{Rzy2016AWPL-RA-Loc} proposed a new single-anchor indoor localization concept based on pattern-RAs. In \cite{Kulas2017AWPL-RA-AOA}, it was demonstrated how a pattern-RA can be used for 2D AoA estimation using only received signal strength (RSS) values. The study in \cite{Kulas2018AWPL-RA-AOA} showed how pattern-RA-enabled AoA estimation in WSN nodes can be improved by applying an interpolation algorithm to radiation patterns recorded during the calibration phase of the AoA estimation process. In \cite{Tarkowski2019AWPL-RA-AOA}, the authors demonstrated that pattern-RA-based AoA estimation in WSNs can be improved by applying a support vector classification approach to RSS values recorded at the antenna’s output port. The work in \cite{Bshara2021Access-RA-AOA} showcased how pattern-RAs can facilitate a rapid optimal beam selection process at mmWave frequencies. Finally, in \cite{Groth2023EuCAP-RA-Loc}, the authors presented improvements to a calibration-free single-anchor indoor localization algorithm designed for BSs equipped with pattern-RAs. Despite these advancements, it is important to note that pattern-RAs are typically limited to discrete beam orientations, resulting in relatively lower angular resolution.  
	
	In contrast, LWAs exploit wave leakage along a guiding structure to generate radiation beams whose direction depends on the operating frequency. This \emph{frequency-dependent} steering allows LWAs to achieve higher angular resolution through continuous frequency scanning, where distinct angles are mapped to specific signal frequencies \cite{Abiel2011TAP-Leaky-AOA,Paaso2013PIMRC-Leaky-AOA,Hakka2014CROWNCOM-Leaky-Loc,Paaso2017TAP-Leaky-AOA,Poveda2019TAP-Leaky-AOA,Ning2020CL-Leaky-AOA,Poveda2020Access-Leaky-AOD,Zaka2023IOTJ-Leaky-AOA-CFO}. The study in \cite{Abiel2011TAP-Leaky-AOA} represents one of the earliest efforts to apply LWAs for AoA estimation, utilizing a composite right/left-handed (CRLH) LWA. In \cite{Paaso2013PIMRC-Leaky-AOA}, the authors proposed a modified unitary MUSIC algorithm for a two-port CRLH LWA. The work in \cite{Hakka2014CROWNCOM-Leaky-Loc} addressed LWA-assisted low-complexity algorithms and evaluated the practical performance of low-complexity cognitive radio primary user (PU) AoA estimation and PU localization using real-world indoor measurement data. In \cite{Paaso2017TAP-Leaky-AOA}, the authors derived the Cramér–Rao bound (CRB) for MUSIC-based AoA estimation with LWAs and presented an extensive performance evaluation of the MUSIC algorithm. In \cite{Poveda2019TAP-Leaky-AOA}, the synthesis of frequency-scanned monopulse radiation patterns using an array of two LWAs was demonstrated, along with a method to estimate the AoA of an incoming beacon signal composed of prescribed tones distributed within the scanning band. In \cite{Ning2020CL-Leaky-AOA}, an improved LWA-enabled AoA estimation method based on the rotational invariance technique was proposed. The authors of \cite{Poveda2020Access-Leaky-AOD} introduced a novel LWA-based Bluetooth beacon, enabling a low-cost direction estimation system. Lastly, \cite{Zaka2023IOTJ-Leaky-AOA-CFO} proposed a joint AoA and carrier frequency offset estimation scheme using LWAs for industrial Internet of Things systems. Compared to pattern-RAs, a key limitation of LWAs worth mentioning is their reliance on large operational bandwidths for high-resolution sensing, which may not be suitable for bandwidth-constrained scenarios.  
	
	Sectorized antennas (SAs) have also been explored in the RA-based sensing literature, abstracting the benefits of reconfigurability regardless of implementation specifics \cite{Werner2013WONS-Sector-Loc,Wang2014WCL-Sector-Loc,Werner2015JSAC-Sector-AOA,Werner2016TVT-Sector-Loc,Pell2022PIMRC-Sector-AOA}. SAs operate by selectively receiving signals from predefined spatial sectors (continuous angular regions) while attenuating out-of-sector signals. Their analog nature ensures that only one sector is active at a time, providing directional precision. In \cite{Werner2013WONS-Sector-Loc}, the authors proposed a low-complexity algorithm for SAs that provides coarse RSS and AoA estimates, deriving asymptotic bounds for its mean square error (MSE) as a function of the antenna parameters. In \cite{Wang2014WCL-Sector-Loc}, the authors presented an analytical performance evaluation of PU RSS/AoA estimation and localization through cooperating cognitive radios. The study in \cite{Werner2015JSAC-Sector-AOA} introduced a high-performance AoA estimator for SAs that does not require cooperation between the transmitter and the localizing network, was broadly applicable to different SA types and signal waveforms, and had low computational complexity. In \cite{Werner2016TVT-Sector-Loc}, the authors derived the CRB on RSS/AoA estimation based on sector powers, studied its asymptotic behavior, and compared the MSE performance of a practical SA-based AoA estimator to the derived CRB. Lastly, \cite{Pell2022PIMRC-Sector-AOA} explored the use of drone formations equipped with SAs to navigate toward a transmitter using AoA, facilitating search-and-rescue applications. Notably, switched-beam antennas (SBAs), which employ analog phase control networks to switch between fixed beam patterns, are occasionally classified as SAs in certain sensing contexts \cite{Gotsis2009TAP-Switch-AOA,Badawy2017WCNC-Switch-AOA}. However, SBAs lack EM domain reconfigurability, as their beam steering relies on phased array techniques rather than dynamic modifications of the antenna’s EM properties. Consequently, SBAs are excluded from discussions of sectorized antennas in RA-based sensing.  
	
	\subsubsection{Extension to IRS-aided Wireless Sensing}
	
	IRS technology has emerged as a promising solution for enhancing wireless sensing capabilities and promoting the development of ISAC systems. By manipulating the propagation environment through reconfigurable metasurfaces, IRS facilitates the creation of virtual LoS paths, thereby addressing inherent challenges in traditional sensing and ISAC scenarios, such as blockages, severe path loss, and limited coverage~\cite{10422881}.
	
	\paragraph{IRS-aided Wireless Sensing}
	The capability of IRS to manipulate EM signals offers a promising new avenue for improving radar sensing performance. In single-target sensing scenarios, the authors in~\cite{9328302} proposed deploying an IRS near the radar receiver to precisely adjust the phase of reflected signals, thus enabling in-phase superposition of echoes and significantly enhancing the received radar power. To address the challenge of radar blind-spot detection, IRS can establish virtual LoS links, thereby enabling effective target detection~\cite{9508883} and accurate parameter estimation~\cite{10138058} in NLoS areas. In multi-target sensing scenarios, the authors in~\cite{9725255} demonstrated that deploying an IRS on the radar side can establish additional echo paths, thereby effectively enhancing the radar's multi-target detection capabilities. However, when multiple targets share the same direct radar-IRS link, it is typically difficult to distinguish between targets in the spatial domain. To address this issue, the authors in~\cite{9938373} proposed a protocol based on time-domain and symbol-domain techniques to distinguish multiple targets, thereby expanding the sensing coverage in multiuser scenarios.
	Nevertheless, in passive IRS-assisted sensing architectures, the sensing signal typically undergoes multiple reflections and suffers from significant path loss, resulting in limited sensing SNR. To mitigate this issue, semi-passive IRS-assisted sensing architectures enable direct reception of echo signals reflected from targets~\cite{10422881,10595504}. As a result, the propagation path is reduced from triple to double reflections, thus enhancing the received SNR and improving sensing accuracy.
	
	From another perspective, conventional IRS-aided sensing systems deploy IRSs as anchor nodes, relying on target-reflected echo signals for detection. However, this approach faces inefficiency when targets exhibit limited radar cross-section (RCS). To overcome this limitation, target-mounted IRS has emerged as a novel paradigm, where IRS is directly integrated onto the target~\cite{10443321,10274514,10437575}. For instance, the authors in~\cite{10274514} considered a target-mounted passive IRS architecture that enables high-precision sensing through tensor-based algorithms, even under limited receiver deployment. In addition, target-mounted IRS offers a novel and cost-efficient solution for enhancing anti-detection capabilities in secure sensing and communication systems~\cite{10839241}. By dynamically reconfiguring the reflection of incident EM waves, a target-mounted IRS can suppress radar echoes to achieve EM stealth or redirect them to generate deceptive signals for spoofing purposes~\cite{10575930,10474137,10919095,10634199}. This enables the target to evade or mislead adversarial detection with high adaptability across temporal, frequency, and spatial domains. Meanwhile, IRS can facilitate covert communications by reinforcing signal strength at legitimate receivers while attenuating or randomizing signal leakage toward potential eavesdroppers~\cite{9496108,9108996}. Compared to conventional anti-detection techniques, target-mounted IRS offers real-time reconfigurability, ultra-low power consumption, and ease of integration, thereby constituting a promising enabler for intelligent anti-detection in highly dynamic wireless environments.

	\paragraph{IRS-aided ISAC}
	As a key technology for future 6G networks, ISAC still faces multi-dimensional technical challenges for effective implementation, particularly suffering from significant performance bottlenecks in signal coverage. To address this issue, IRS can flexibly reconfigure wireless channels, thereby effectively extending both sensing and communication coverage. In recent years, considerable efforts have also been devoted to integrating IRS with ISAC systems~\cite{10243495,9782100,10254508}. In general, existing works on IRS-aided ISAC have focused on two main application scenarios. In the first scenario, IRS is primarily used to enhance communication performance, while direct transceiver-target links are used for sensing~\cite{9909807,9913311,10086570}. Specifically, the authors in~\cite{9909807} concentrated on designing the transmit/receive beamforming matrices and the phase shifts of passive IRS in multiuser scenarios. The authors in~\cite{9913311} aimed to minimize the transmit power at the BS by jointly optimizing active and passive beamforming in the presence of interference introduced by the IRS. Meanwhile, the authors in~\cite{10086570} addressed a sum-rate maximization problem, subject to constraints on target AoA estimation performance. The second scenario primarily aims at improving the sensing performance, by establishing a virtual LoS link between the BS and the target~\cite{9771801,9364358,10138058}. For example, the authors in~\cite{9771801} optimized the radar beampattern to concentrate more transmit power toward the target, thereby enhancing sensing performance. Moreover, the authors in~\cite{9364358} analyzed the received SINR at the BS with IRS passive sensing, while the authors in~\cite{10138058} considered the same scenario by formulating a CRB minimization problem to enhance sensing performance. Furthermore, it is worth noting that a critical advantage of IRS in ISAC is its capability to establish virtual LoS paths, simultaneously enhancing both communication reliability and sensing accuracy. By intelligently optimizing IRS reflection patterns, an enhanced balance can be achieved between communication and sensing performance~\cite{10464353,9898900,10304548}.
	
	In summary, IRS technology has a profound impact on wireless sensing and ISAC systems by intelligently manipulating the propagation environment, thereby overcoming traditional limitations. Continued research into IRS deployment strategies, reflection design optimization, and waveform design will be essential to fully realizing the potential of IRS-aided wireless sensing and IRS-aided ISAC, ultimately paving the way for efficient and effective next-generation wireless networks.
	
	\begin{figure*}[!t]
		\centering
		\includegraphics[width=140mm]{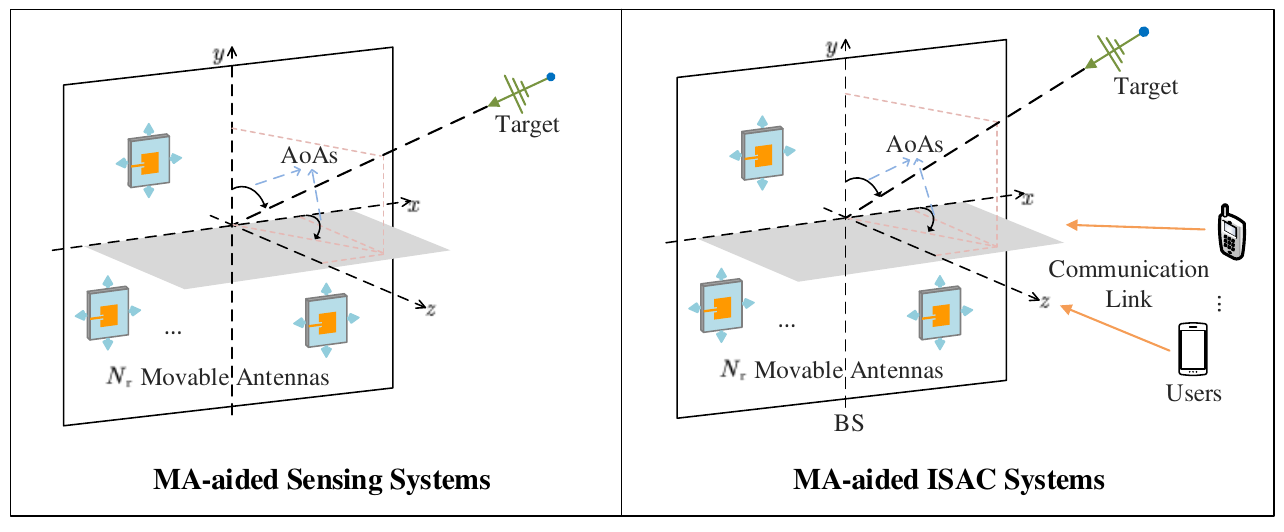}
		\caption{Illustration of MA-aided sensing and ISAC systems.}
		\label{Fig_sensing}
	\end{figure*}

	\subsection{MA for Wireless Sensing and ISAC}
	\label{subsec:sensing_ma}
	As shown in Fig.~\ref{Fig_sensing}, MAs/6DMAs offer significant potential not only for wireless communications but also for enhancing wireless sensing capabilities and enabling more effective ISAC systems. By providing the ability to dynamically optimize antenna positions and array geometries, MAs/6DMAs unlock new DoFs to improve sensing accuracy, resolution, and the synergy between sensing and communication functions \cite{zeng2024fixed}.

	\subsubsection{Sensing}
	The key advantage of MAs in wireless sensing lies in their ability to enlarge the effective antenna aperture for forming narrow sensing beams, as well as their capability to adapt antenna positions in response to environmental variations \cite{ma2024MAsensing, Chen2024moving, shao2024exploiting}. This can lead to significant improvements in various sensing applications. Traditional mechanically rotatable radar systems, which scan sensing targets by rotating a directional antenna, represent an early form of antenna movement for sensing. MAs generalize this concept by enabling full 3D translational and rotational control of antenna elements, allowing for more flexible scanning patterns and enhanced adaptability to complex and dynamic environments. Besides, a closely related concept is SAR \cite{Moreira2013SAR}, where the movement of antennas over time is used to form a large aperture for high-resolution imaging. {In contrast, MAs offer the potential of more adaptive aperture synthesis, supporting flexible 1D/2D/3D movements and real-time trajectory optimization \cite{ma2025movabletra}.}
	
	For target detection, localization, and tracking, MAs can significantly enhance performance in both far-field and near-field scenarios. In far-field scenarios, MAs can enhance sensing performance by enlarging the effective array aperture, which is critical for improving angular resolution and the accuracy of AoA estimation \cite{ma2024MAsensing, mao2025movable}. In near-field scenarios, where targets are close to the antenna array, the wavefronts are spherical and carry both range and angle information. By optimally sampling spherical wavefronts, MA systems can accurately estimate both range and angle, resulting in superior localization precision compared to conventional FPA arrays \cite{Chen2024moving}. The ability of MAs to dynamically tailor their geometry enables effective exploitation of the distinct propagation characteristics in both far-field and near-field scenarios for a wide range of sensing objectives.
	
	Several design issues are critical for MA-aided sensing. First, sufficient time and an adequate movement region are required for the antennas to transmit and receive sensing echos as well as synthesize a desired array aperture, especially when sensing static or slowly varying environments/targets. The size of the movement region directly determines the achievable resolution. For dynamic targets, the movement speed and the system’s ability to predict and track target motion become essential. Moreover, the geometry of the MA array should be carefully designed to avoid grating lobes, which can lead to ambiguities in AoA estimation \cite{ma2024MAsensing}. Lastly, accurate calibration is important to enable coherent signal processing and ensure high sensing accuracy.
	
	\subsubsection{ISAC}
	ISAC enables the simultaneous sensing and communication by sharing hardware and/or radio resources \cite{Shao2022target}. Given the advantages of MAs in enhancing spatial multiplexing and beamforming in wireless communication, as well as improving spatial resolution in wireless sensing, MA is a promising technology for ISAC applications to achieve increased design flexibility and enable combined performance gains \cite{kuang2024movableISAC, ma2025MAISAC, li2024MAISACMag, WuHS_MA_RIS_ISAC, lyu2024flexibleISAC, peng2024jointISAC, wang2024multiuser}. The key advantage of MAs in ISAC systems lies in their ability to dynamically adjust antenna positions, enabling flexible trade-offs or simultaneous enhancement of both sensing and communication performance \cite{kuang2024movableISAC, ma2025MAISAC, li2024MAISACMag, WuHS_MA_RIS_ISAC, lyu2024flexibleISAC, peng2024jointISAC, wang2024multiuser, guo2024movable, MaY2024movableISAC, zhou2024fluidISAC}. For instance, an MA array can expand its aperture to enhance sensing resolution for target detection, while optimizing the array geometry based on instantaneous or statistical CSI to enable efficient beamforming and interference mitigation in multiuser communication. MA-aided ISAC systems can adapt more effectively to varying system requirements and environmental conditions compared to FPA-based systems.
	
	However, the integration of MAs into ISAC systems introduces unique design challenges. A fundamental issue is the potentially conflicting requirements between sensing and communication \cite{ma2024MAsensing}. For example, very sparse antenna arrays that support high spatial multiplexing in multiuser communication can also introduce grating lobes, which cause ambiguities in sensing tasks such as target detection and localization. This trade-off necessitates the joint optimization of antenna positions, communication beamformers, and sensing waveforms to effectively balance or improve both sensing and communication performance. Additionally, the allocation of MA movement time and energy between sensing and communication must be carefully managed, especially when the two tasks operate on different timescales or have differing priorities. These joint optimization problems are typically high-dimensional and non-convex, demanding efficient and real-time algorithms for adapting to dynamic environments \cite{ma2025MAISAC}. Moreover, the real-time antenna position optimization to support both sensing and communication tasks requires robust control strategies and often predictive capabilities to overcome movement latency \cite{ma2025MAISAC}. {To reduce antenna movement complexity, a promising research direction to MA-aided ISAC is dynamically reconfiguring the MA array geometry. For example, to manage the fundamental trade-off between spatial multiplexing and grating lobe suppression, we can synthesize a dense array geometry during sensing intervals to ensure high-precision sensing without ambiguities. Subsequently, the MA array can form a sparse geometry during communication intervals to maximize the spatial multiplexing gain for multiuser communication. Such array-sparsity adaptation allows ISAC systems to easily optimize their configuration for specific sensing or communication tasks.} Finally, standardization is needed for MA-aided ISAC, including protocols for movement scheduling as well as joint waveform design, all of which are critical for practical deployment. 
	
	\subsection{Performance Comparison}
	\label{subsec:sensing_comparison}
	\begin{figure}[!t]
		\centering
		\includegraphics[width=80mm]{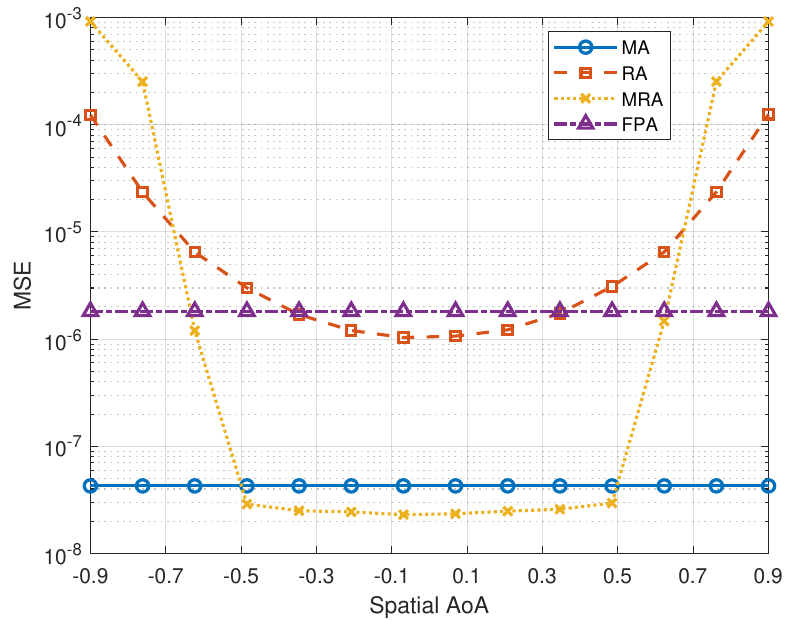}
		\caption{Comparison of AoA estimation MSE for MA and RA systems versus the true spatial AoA.}
		\label{Fig_MA_RA_sensing}
	\end{figure}
	
	To evaluate and compare the performance of MA-aided and RA-aided wireless sensing systems, Fig.~\ref{Fig_MA_RA_sensing} presents the MSE of AoA estimation versus the true spatial AoA. A linear array consisting of $18$ antennas is considered, and the MUSIC algorithm is employed for AoA estimation. The receive SNR is fixed at $5$ dB. Four antenna configurations are considered: (i) MA: The antennas are divided into two equal groups, with one half placed at the leftmost end and the other at the rightmost end of the array segment, each forming a subarray with $\lambda/2$ inter-antenna spacing; (ii) RA: A ULA with $\lambda/2$ spacing, where the antennas are partitioned into three equal groups, each with a distinct directional radiation pattern \cite{wang2025EMally}; (iii) MRA: This configuration combines the antenna placement of the MA scheme with the directional radiation patterns of the RA scheme; (iv) FPA: A ULA with $\lambda/2$ inter-antenna spacing and isotropic radiation patterns.
	
	As illustrated in Fig.~\ref{Fig_MA_RA_sensing}, the MA scheme consistently achieves lower MSE than the FPA scheme across the full AoA range. This is because by enlarging the effective aperture, the MA scheme can achieve higher angular resolution for AoA estimation than the FPA scheme. {The RA and MRA schemes exhibit a U-shaped MSE profile, i.e., achieving lower MSE near the array boresight (i.e., spatial AoA of $0$) but higher MSE at wider angles. This is because the three reconfigurable patterns are directed toward spatial AoAs of $-0.5$, $0$, and $0.5$, thereby enhancing the effective SNR for targets near these directions while reducing it elsewhere \cite{wang2025EMally}.} The MRA scheme, which integrates the large aperture of the MA scheme with the directional focusing capability of the RA scheme, achieves the best performance near the center of the angular range, even surpassing the MA scheme. This shows the benefit of jointly leveraging aperture gain and energy focusing. These observations reveal a fundamental trade-off in array design for sensing applications: while a larger aperture ensures higher angular resolution over a wide AoA range, reconfigurable radiation patterns can further enhance performance within specific angular sectors of interest.
	
	\textit{Lessons Learned}: This section demonstrates that RA and MA provide distinct advantages for sensing applications. RA enables a single element to achieve angular resolution by reconfiguring its radiation pattern, whereas MA synthesizes a larger effective aperture through physical movement. This leads to a fundamental trade-off: MA offers consistently high angular resolution across a wide angle, while RA achieves superior estimation accuracy within specific angular sectors by focusing energy. Moreover, a hybrid MRA system that integrates and jointly optimizes both functionalities could achieve unparalleled performance, combining the wide-angle and high-resolution capability of MA with the ability of RA to enhance sensing accuracy in selected angular sectors.
	
	\section{Future Works}
	\label{sec:future}
	The development of RA and MA/6DMA is promising for enhancing wireless communication and sensing systems. However, to fully unlock their potential and enable widespread practical deployment in future wireless networks such as 6G, several research challenges and opportunities need to be addressed. This section outlines key directions for future investigation.
	
	\subsection{Efficient Antenna Architectures}
	To enable the practical implementation of RA and MA systems, efficient antenna architectures are essential to enhance system performance, reduce cost, and support broader deployment. Future research in RA architectures should focus on developing novel tunable materials that provide faster switching speeds, wider reconfiguration ranges, lower losses, and reduced power consumption compared to current technologies such as liquid crystals and ferrites. Advances in MEMS and nano-electromechanical system (NEMS) could enable more compact and efficient reconfiguration mechanisms. Additionally, the development of multi-functional RAs capable of simultaneously adjusting multiple parameters is a promising research direction.
	
	For MA, the development of more compact, energy-efficient, and faster mechanical actuation technologies is essential. This includes research into smart materials (e.g., shape memory alloys and piezoelectric actuators) and advanced robotic systems that can provide precise multi-dimensional movement with minimal overhead. A key focus is creating scalable architectures, such as movable subarrays or cross-linked MA arrays \cite{zhu2025multiuser}, which aim to achieve near-optimal communication and sensing performance with the least possible movement complexity. Additionally, developing lightweight, robust, and easily deployable MA arrays leveraging origami-inspired foldable structures or inflatable technologies is important for their integration into diverse platforms,  such as mobile devices, UAVs, and BSs.
	
	A particularly promising direction is the development of integrated RA and MA architectures. In such hybrid systems, each RA element or the RA array can be physically moved, combining the rapid, fine-grained reconfigurability of RAs with the slower but larger-scale spatial optimization of MAs. However, realizing such integrated systems will involve significant challenges, including the co-design of reconfiguration and movement mechanisms, mitigation of mutual interference, assurance of energy efficiency, and the development of hierarchical control frameworks to coordinate their operation. {A practically critical issue is the management of dynamic EM coupling, as changes in antenna states or positions modify the mutual coupling between antennas. This gives rise to a complex and high-dimensional EM problem that must be addressed in both hardware design and signal processing to ensure system stability and performance in dense hybrid arrays.} {Moreover, in hybrid RA/MA architectures, an important open challenge lies in the design of hierarchical and cross-layer control frameworks. In particular, it remains unclear how to jointly control the MA and RA layers. For example, large-scale statistical CSI available at the MA layer may be leveraged to guide or constrain the adaptation space of the RA configurations, while the rapid state changes of the RA may need to be abstracted or decoupled from the MA movement optimization to avoid excessive signaling and control overhead. Addressing such cross-layer control and information-coupling issues is a promising direction for future research.}
	
	\subsection{Channel Estimation/Acquisition for RA and MA Systems}
	Accurate and efficient channel knowledge is essential to fully exploit the DoFs offered by RA and MA systems. For RA systems, future work should focus on fast channel estimation techniques tailored to their specific reconfigurable parameters. This includes methods for efficiently learning how the channel responds to variations in antenna parameters, such as beam patterns or polarization states, potentially leveraging compressed sensing or AI-driven approaches to reduce the training overhead associated with numerous RA configurations.
	
	For MA systems, although both model-based and model-free channel acquisition methods have been explored, significant challenges remain \cite{ma2023MAestimation, xiao2023channel, zhang2024TensorCE, Skouroumounis2023fluidCE, zhang2023successive, ji2024correlation, new2024CE, cui2024near, zhang2024MLCE}. In particular, reducing the overhead of channel acquisition over large or high-dimensional movement regions is critical \cite{ma2023MAestimation, xiao2023channel}. Robust acquisition methods that can tolerate positioning inaccuracies or operate during continuous movement are essential. {Furthermore, considering that model-based algorithms are sensitive to MAs' positions and orientations, it is crucial to develop schemes that are resilient to positioning and orientation inaccuracies, or those that jointly estimate these inaccuracies alongside the channel parameters to mitigate calibration-related performance degradation.} Moreover, statistical channel acquisition techniques that capture long-term spatial characteristics without requiring exhaustive instantaneous measurements are important for reducing the overhead of frequent antenna movement \cite{chen2023joint}. Channel acquisition with antenna rotation also needs further investigation by exploiting directional sparsity \cite{shao2024distributed,shao2025hybridnear}.
	
	For integrated RA and MA systems, channel acquisition becomes exceptionally challenging due to the high-dimensional channel space, which varies with both physical position/orientation and the internal antenna state. Developing novel frameworks that can efficiently represent and estimate such high-dimensional and configuration-dependent channels remains a key open research direction.
	
	\subsection{Low-complexity Antenna Configuration/Movement Strategies}
	Optimizing the state of RAs or the position/orientation of MAs can be computationally intensive, necessitating the development of low-complexity strategies suitable for real-time implementation. For RA systems, it is important to develop efficient algorithms to select optimal antenna configurations based on limited channel acquisition or learned environment-specific policies. AI-based techniques are promising for enabling RA to autonomously adapt its states with reduced computational overhead and without requiring full CSI.
	
	For MA systems, it is important to acquire antenna positions using low-overhead channel acquisition techniques and low-complexity antenna position optimization methods \cite{zeng2024csi}. Future research should focus on efficient algorithms that can obtain effective MA positions with minimal time overhead. A crucial challenge lies in managing movement overhead \cite{Vahid2024movable, Vahid2024MAplan, wang2024MAdelay}, where intelligent control strategies must determine how to move antennas by explicitly balancing the trade-offs between performance gains and the associated costs in energy consumption, latency, and computational complexity \cite{li2024minimizing}. {Furthermore, optimization strategies should account for the physical dynamics of the motors and actuators. Practical mechanical constraints such as inertia, settling time, friction, and gear backlash are critical factors that affect the feasibility of rapid and precise antenna movement/rotation. Therefore, future research should develop motion-aware optimization frameworks that integrate these mechanical realities into the antenna movement/rotation scheduling process.} Trajectory optimization of multiple MAs without antenna coupling is another critical area \cite{li2024minimizing}. Additionally, tailored optimization frameworks for MAs with constrained mobility (e.g., linear sliding arrays or group-based MAs \cite{ning2024movable, yichi2024movable, lu2024group, shi2024capacity}) can reduce control complexity while still achieving significant performance gains.
	
	For integrated RA and MA systems, hierarchical control and optimization strategies are promising for reducing control complexity. Such strategies may involve optimizing MA positions based on large-scale or statistical channel characteristics over longer timescales, while RA configurations are adjusted more rapidly in response to instantaneous or local channel conditions.
	
	\subsection{Synergy with Other Technologies and Applications}
	The full potential of RA and MA systems can be realized through their synergy with other emerging wireless technologies and applications.
	
	\subsubsection{Mobile Edge and Over-the-Air Computing}
	The interaction between RA/MA and mobile edge computing (MEC) is particularly promising \cite{xiu2024delayMAMEC, xiu2024latencyMAMEC, zuo2024fluid, ChenPC_MA_WPT_MEC}. MAs can dynamically optimize links to edge servers for efficient computation offloading, while RAs can adapt beam patterns to support mobile users accessing MEC services. Another related area is over-the-air computation (AirComp), where MAs can reshape wireless channels to improve the accuracy of distributed data aggregation at a sink node \cite{zhang2024AriComp, li2024over, cheng2023movableAirComp}.
	
	\subsubsection{Physical Layer Security (PLS)}
	RAs/MAs offer new opportunities for PLS by reconfiguring channels to enhance the link to a legitimate receiver while simultaneously degrading the link to an eavesdropper \cite{ma2025movablemag,zhu2025towed,hu2024secure, cheng2024secure, tang2024secure, DingJZ_MA_FD_secure_1, liao2024MAjamming}. By jointly optimizing beamformers and antenna states/positions, it is possible to improve secrecy rates, even when the eavesdropper's CSI is unavailable \cite{HuGJ_secure_CSIfree_MA, FengZY_MA_secure, cheng2024MAnoCSI}, or to enhance covert communications \cite{mao2024MAcovert, wang2024MAcovert, Xie2024MARIScovert, Liu2024MAcovert}.
	
	\subsubsection{Wireless Power Transfer (WPT)}
	In WPT and simultaneous wireless information and power transfer (SWIPT) systems, RAs/MAs can improve energy transmission efficiency \cite{ZhangL_FAS_WPT, zhou2024fluidISAC, XiaoJH_MA_RIS_WPT, Farshad_FAS_RIS}. Antenna states/positions can be optimized to balance the trade-off between maximizing the SINR for information decoding and maximizing the received power for energy harvesting \cite{Psomas_WPT_survey, WongKK_FAS2, Christodoulos_FAS_SWIPT_1, LaiXZ_FAS_WPT, Christodoulos_FAS_SWIPT_2, LinX_FAS_WPT_1, LinX_FAS_WPT_2}. {An interesting direction is the potential of MA systems to harvest energy from their own mechanical movement by integrating piezoelectric or kinetic harvesters into the actuation mechanism. This harvested energy could potentially power low-energy RA reconfiguration circuits or other low-power electronics, thereby creating more self-sustaining and autonomous wireless systems.}
	
	\subsubsection{Next-Generation Multiple Access (NGMA)}
	NGMA schemes, such as rate-splitting multiple access (RSMA) \cite{Mao2022RSMA} and NOMA \cite{liu2022NGMA, Clerckx2024multiple}, can benefit from RA/MA technology. By adjusting antenna states/positions to create favorable channel conditions that either reduce inter-user correlation for spatial division multiple access or increase it for selected user subsets to improve the performance of NOMA and RSMA \cite{Li2024MAupNOMA, Zhou2024MAdownNOMA, He2024MANOMA, gao2024MApowerNOMA, amhaz2024optimizing, xiao2024movableNOMA, zhang2024movableRSAM, ghadi2024fluid}.
	
	
	\subsubsection{AI and Large Language Models (LLMs)}
	AI techniques are crucial for managing the complexity of RA/MA systems \cite{zhu2023MAMag, WangC_FAS_RIS_AI_survey, Kang2024DeepMA, tang2024deepMA, weng2024learnMA, Waqar2024leaningfluid, bai2024Deepmovable, zhao2024FLmovable, ahmadzadeh2024enhancement,li2025ai}. The emergence of LLMs offers new possibilities for generating adaptive control policies for RA state selection and MA movement in real-time, responding to dynamic environments and system requirements \cite{Creswell2018GAN, jiang2024LLM}. {For example, an LLM could process a high-level natural language command and subsequently translate this intent into a multi-variable optimization policy for the joint control of RA states and MA positions. This enables the MA/RA system to respond dynamically to complex environmental conditions and human-centric objectives without relying on manual configurations.}
	
	\section{Conclusions}
	\label{sec:conclusion}
	This paper has provided a comprehensive survey on the fundamentals, architectures, and applications of RA and MA technologies. We began by reviewing the historical development of both RA and MA technologies, tracing their parallel evolution and highlighting their promising application scenarios in future wireless networks. We then presented a detailed overview of the hardware architectures for both RAs and MAs, covering their classification, implementation methods, and a comparative analysis of their distinct mechanisms and performance metrics. Subsequently, we focused on the application of RAs and MAs in wireless communications, examining their respective performance benefits and design methodologies. The discussion was then extended to wireless sensing and ISAC, reviewing RA- and MA-enabled techniques and their unique advantages. We also presented numerical performance comparisons to illustrate the distinct and complementary characteristics of RA and MA systems in various scenarios. Finally, we outlined key challenges and identified promising future research directions. As the exploration of RA and MA technologies is still in its early stage, we hope this survey will serve as a valuable resource for researchers and practitioners, inspiring further innovations to unlock the full potential of these promising technologies in realizing intelligent and adaptive wireless networks.
	
	\bibliographystyle{IEEEtran}
	\bibliography{IEEEabrv,IEEEexample}
	
\end{document}